\documentstyle[prb,eqsecnum,aps,epsf,ifthen,multicol]{revtex}

\tighten
\newcommand{\wide}[2]{
\end{multicols}
\widetext
\noindent
\ifthenelse{\equal{#1}{t}}
{}
{
\raisebox{0.1in}[0in][0.02in]{$\rule{3.575in}{0.002in}
\rule{0.002in}{0.08in}$}
}
#2
\ifthenelse{\equal{#1}{b}}
{}
{
{\raisebox{-0.1in}[0in][0.02in]
{\hspace{3.575in}$\rule{0.002in}{0.08in}
\rule[0.08in]{3.575in}{0.002in}$}
}
}
\begin{multicols}{2}
\noindent
}

\begin{document} \draft \title{Coulomb blockade in
metallic grains at large conductance} \author{I.S. Beloborodov and A.V.
Andreev} \address {Bell Laboratories, Lucent Technologies, Murray
Hill, NJ 07974\\
Department of Physics, University of
Colorado, CB 390, Boulder, CO 80390 \\
Theoretische Physik III, Ruhr-Universit\"{a}t Bochum, 44780 Bochum,
Germany} \date{\today} \maketitle

\begin{abstract}
  We study Coulomb blockade effects in the thermodynamic quantities of a
  weakly disordered metallic grain coupled to a metallic lead by a
  tunneling contact with a large conductance $g_T$.  We consider the case
  of broken time-reversal symmetry and obtain expressions for both the
  {\em ensemble averaged} amplitude of the Coulomb blockade oscillations
  of the thermodynamic potential and the correlator of its {\em mesoscopic
    fluctuations} for a finite mean level spacing $\delta$ in the grain.
  We develop a novel method which allows for an exact evaluation of the
  functional integral arising from disorder averaging.  The results and
  the method are applicable in the temperature range $\delta \ll T \ll
  E_C$.
\end{abstract}

\begin{multicols}{2}

\section{Introduction}

\begin{figure}
  \epsfysize =4cm \centerline{\epsfbox{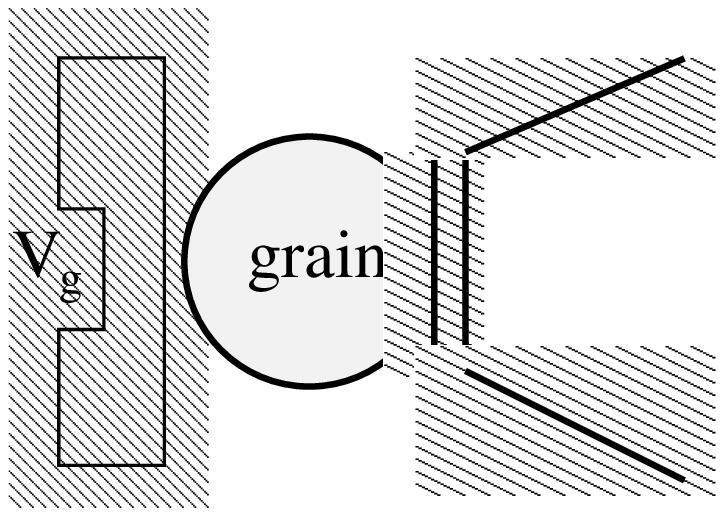}}
\caption{Schematic drawing of a disordered metallic grain coupled by a
  tunneling contact to a metallic lead. The charge of the grain is
  controlled by the gate voltage $V_g$}
\label{dot}
\end{figure}

The study of electron-electron interactions in mesoscopic systems has been
at the focus of experimental and theoretical interest over the past two
decades.  One of the most striking manifestations of electron interactions
at low temperatures is the phenomenon of Coulomb
blockade.~\cite{Kouwenhoven} It can be observed by measuring, say the
charge of a metallic grain which is connected by a tunneling contact to a
metallic lead and is capacitively coupled to a metallic gate that is
maintained at voltage $V_g$, as in Fig.~\ref{dot}.

For a poorly conducting contact the charge in the grain is quantized at
low temperatures and exhibits a characteristic step-like dependence on the
gate voltage $V_g$. This leads to oscillatory gate voltage dependence of
all physical quantities, referred to as the Coulomb blockade oscillations.
The amplitude of these oscillations decreases with increasing transparency
of the contact and depends not only on the total conductance of the
contact but on the individual transparencies of the tunneling channels.

In the case of a few transmission channels in the contact and in the limit
of vanishing mean level spacing, $\delta \to 0$, the Coulomb blockade
oscillations of the thermodynamic, transport and thermoelectric quantities
are completely smeared at perfect
transmission.~\cite{Matveev95,Furusaki95,Andreev01} For a finite mean
level spacing in the grain, the Coulomb blockade oscillations remain
finite even at perfect transmission and exhibit strong mesoscopic
fluctuations.~\cite{Aleiner,Aleiner_review}

In the case of a contact with many weakly transmitting tunneling channels,
the Coulomb blockade oscillations at large conductance have been studied
in the limit of vanishing mean level spacing in the grain, $\delta$, and
shown to be exponentially
suppressed~\cite{Zaikin91,Schoen,Flensberg,Grabert,Hofstetter} $\sim \exp
(-g_T/2)$.  Here $g_T$ is the conductance of the tunneling contact
measured in units of $e^2/(2\pi \hbar)$.  In the $\delta \to 0$ limit for
a non-random contact the mesoscopic fluctuations in the Coulomb blockade
oscillations vanish.  At finite mean level spacing the mesoscopic
fluctuations in the thermodynamic quantities for a weakly transmitting
tunneling contact were studied in Ref.~\onlinecite{Kaminski} and for a
multi-channel contact near perfect transmission in
Ref.~\onlinecite{Aleiner_review}.  In the case of a random diffusive
contact and vanishing mean level spacing in the grain the Coulomb blockade
oscillations were studied in Refs.~\onlinecite{Nazarov,Kamenev}.

Here we study the Coulomb blockade oscillations in the thermodynamic
quantities of a disordered metallic grain with finite mean level spacing
$\delta$ which is connected to a metallic lead by a non-random
multi-channel tunneling contact and is capacitively coupled to a gate, as
in Fig.~\ref{dot}.  The conductance of the tunneling contact is assumed to
be large, $g_T \gg 1$.  We assume that electron-electron interactions in
the metallic lead may be neglected due to screening and treat the Coulomb
interaction of electrons in the grain in the framework of the constant
interaction model:
\begin{equation}
  \label{eq:cint}
  \hat{H}_C=E_C\left(\hat{N}-q\right)^2.
\end{equation}
Here $E_C$ is the charging energy, $\hat{N}$ is the operator of the number
of electrons in the grain, and $q$ is the dimensionless parameter which is
proportional to the gate voltage $V_g$ and has the meaning the number of
electrons that minimizes the electrostatic energy of the grain.  We
consider a disordered metallic grain in which electrons move in the
presence of a random impurity potential and study both ensemble averaged
thermodynamics quantities of the grain and their mesoscopic fluctuations.
We assume that the Thouless energy of the grain, $E_T=D/L^2$, where $D$ is
the diffusion coefficient and $L$ is the grain size, is greater than the
charging energy, $E_T >> E_C$. In addition, we assume that the time
reversal symmetry is broken by a magnetic field $H$ and that the cooperon
gap $DeH/(\hbar c)$ is greater than the charging energy $E_C$.  Under
these conditions the ensemble of disorder potentials is equivalent to the
unitary random matrix ensemble.~\cite{Efetov97}

We use the replica formalism to treat disorder averaging in the presence
of interactions.~\cite{Fin} The application of the
Itzykson-Zuber~\cite{Harish-Chandra57,Zuber} integral enables us to
evaluate the resulting functional integral over the $Q$-matrix exactly.
The method used here may have useful applications to the treatment of
non-perturbative interaction effects in granulated disordered
systems.~\cite{Beloborodov}

The main results of the paper are the analytic expressions for the average
oscillatory part of the grand canonical potential, Eq.~(\ref{2}) and for
the oscillatory part of the correlator of the grand canonical potentials
at different values of the gate voltage,
Eq.~(\ref{eq:connectcorrresult}).

The paper is organized as follows: In section \ref{noninteract} we
introduce the formalism by considering a closed disordered metallic grain
with non-interacting electrons in the zero dimensional limit, and obtain
the ensemble average and the fluctuations of the thermodynamic potential
$\Omega(q,T)$.  In section \ref{interaction} we apply this method to
compute the average and the mesoscopic fluctuations of the amplitude of
the Coulomb blockade oscillations in the thermodynamic potential
$\Omega(q,T)$.  We discuss our results in section \ref{sec:conclusion}.

\section{Thermodynamics of non-interacting electrons}
\label{noninteract}

First, to illustrate the formalism we consider the thermodynamic
quantities of an isolated disordered metallic grain. In this section we
neglect the electron-electron interactions in the grain. For simplicity,
in this section we consider spinless electrons.

We are working within the grand canonical ensemble with the chemical
potential $\mu$. All thermodynamic quantities can be obtained from the
grand canonical potential $\Omega_0 (\mu, T) =-T \ln Z_0(\mu,T) $.  Here
the subscript $0$ indicates that the quantities pertain to non-interacting
electrons and $Z_0(\mu,T)$ denotes the partition function which can be
written as a functional integral over the fermionic variables $\psi$,
$\bar{\psi}$ as \wide{m}{
\begin{equation}
Z_0(\mu,T)=\int\prod\limits_{x=1}^{N}\prod\limits_{n}
d\bar{\psi}_{x,n}d{\psi_{x,n}}
\exp\left[\sum\limits_{n,m}\sum\limits_{x,y=1}^N
\bar{\psi}_{x,n}\left(i \hat{\varepsilon}
- \delta_{nm}(\hat{H}_{0,xy}-\mu \delta_{xy})
\right)\psi_{y,m} \right].
\label{Z}
\end{equation}
} In this equation the fermion variables are labeled by the Hilbert space
indices $x$ and $y$ and by the Matsubara indices $n$ and $m$, the diagonal
matrix $\hat{\varepsilon}=\delta_{xy}\delta_{nm}\varepsilon_n$ is
expressed through the fermionic Matsubara frequencies
$\varepsilon_n=(2n+1)\pi T$, $\hat{ H}_0 $ is the single particle
Hamiltonian, and $\mu$ is the chemical potential.

We consider the case of broken time-reversal symmetry. Therefore, in the
zero dimensional limit~\cite{Efetov97} the single particle Hamiltonian
$\hat H_0$ can be modeled by a random $N\times N$ Hermitian matrix
belonging to the Gaussian Unitary Ensemble (GUE) with the probability
distribution
\begin{equation}
P(H_0)\sim \exp\left(-\frac{N}{2}\rm Tr \hat{H}_0^2\right).
\label{distribution}
\end{equation}
As is well known~\cite{Mehta}, the average single particle density of
states (DOS) $\nu(E)$ for such a distribution is given by the semi-circle
law
\begin{equation}
\nu(E)=\frac{N}{\pi}\sqrt{1-\frac{E^2}{4}}.
\label{dos}
\end{equation}
Note we normalized the distribution (\ref{distribution}) in such a way
that the width of the semi-circle is equal to unity. Thus, throughout this
paper all energies are measured in units of the width of the semi-circle.
For simplicity, we assume that the chemical potential $\mu$ lies near the
middle of the semi-circle, where the mean level spacing in the dot is
given by
\begin{equation}
\label{eq:delta}
\delta=\frac{\pi}{N}.
\end{equation}

Below we will consider the {\em ensemble averaged} thermodynamic
properties of the dot. For this purpose we resort to the replica
trick~\cite{Edwards75} and find the averaged replicated partition function
\begin{equation}
  \label{eq:zrepl}
  \langle  Z^\alpha _0(\mu) \rangle = \left\langle \exp \left(-\alpha
  \frac{\Omega_0(\mu)}{T}\right)  \right\rangle,
\end{equation}
where $\langle \ldots \rangle$ denotes the averaging over the ensemble of
random matrices $\hat H_0$ with the probability distribution function
defined in Eq.~(\ref{distribution}), and $\alpha$ is the number of
replicas.  The function $\langle Z^\alpha _0(\mu) \rangle$ is initially
determined for the positive integer number of replicas $\alpha$ and then
analytically continued to $\alpha \to 0$.  The $n$-th cumulant of the
thermodynamic potential $\langle \langle \Omega_0(\mu)^n \rangle \rangle$
with respect to the distribution Eq.~(\ref{distribution}) is then found
from $\langle Z^\alpha _0(\mu) \rangle$, Eq.~(\ref{eq:zrepl}) by using the
formula
\begin{equation}
  \label{eq:cumulants}
  \langle \langle \Omega_0(\mu)^n \rangle \rangle=T^n\left. \frac{d^n\ln
      \langle  Z^\alpha _0(\mu) \rangle}{(-1)^n d
      \alpha^n}\right|_{\alpha=0}.
\end{equation}

We express the replicated partition function $Z^\alpha_0(\mu)$ through the
functional integral over the replicated fermionic fields $\psi^j$ as in
Eq.~(\ref{Z}) and average $ Z_0^{\alpha}(\mu)$ with respect to the
probability distribution function Eq.~(\ref{distribution}). As a result we
obtain \wide{m}{
\begin{eqnarray}
\label{Z2}
\langle Z^\alpha_0(\mu)\rangle&=&\int\prod\limits_{j=1}^{\alpha}
\prod\limits_{x=1}^{N}\prod\limits_{n}
d\bar{\psi}_{x,n}^jd{\psi_{x,n}^j}
\exp\left[\sum\limits_{n}\sum\limits_{x=1}^N
\sum\limits_{j=1}^{\alpha}\bar{\psi}_{x,n}^j\left( i\varepsilon_n
+\mu  \right)\psi_{x,n}^j
+\frac{1}{2N}\sum\limits_{x,y=1}^N
\left|\sum\limits_{n}\sum\limits_{j=1}^{\alpha}\bar{\psi}_{x,n}^j
\psi_{y,n}^j\right|^2\right] .
\end{eqnarray}
}

Next, we introduce a Hermitian $\alpha 2M \times \alpha 2M$ matrix
$\hat{Q}$ to decouple the quartic term in Eq. (\ref{Z2}) via the
Hubbard-Stratonovich transformation: \wide{m}{
\begin{eqnarray}
  \label{eq:hubstrat}
\exp\left[  \frac{1}{2N}\sum\limits_{x,y=1}^N
\left|\sum\limits_{n}\sum\limits_{j=1}^{\alpha}\bar{\psi}_{x,n}^j
\psi_{y,n}^j\right|^2\right]=c_{\alpha\times 2M}\int
d[\hat{Q}]\exp\left\{-\frac{N}{2}\rm Tr\hat{Q}^2+ i
\sum\limits_{x=1}^N
\sum\limits_{n,m}\sum\limits_{i,j=1}^{\alpha}\bar{\psi}_{x,n}^i
Q^{ij}_{nm}
\psi_{x,n}^j \right\},
\end{eqnarray}
} where
\begin{equation}
c_{\alpha\times
2M}=\left(\frac{N}{2\pi}\right)^{(\alpha2M)^2/2}
2^{\alpha M(\alpha2M-1)},
\label{const}
\end{equation}
and the trace in Eq.~(\ref{eq:hubstrat}) is taken over both the Matsubara
and the replica indices. Here $2M$ is the number of Matsubara frequencies
in each replica. We keep it finite at this point to make the sums over the
Matsubara frequencies well defined. Eventually the limit
$2M\rightarrow\infty$ will be taken to obtain the final expressions. The
elements of the $\hat{Q}$-matrix are labeled by four indices:
$Q_{nm}^{ij}$, two of which refer to the replica space $i,j$, and two
others, $n,m$ refer to the Matsubara space.  After the integration over
the fermionic variables in Eq.~(\ref{eq:hubstrat}) we obtain \wide{m}{
\begin{equation}
\langle Z^{\alpha}_0 (\mu)\rangle=c_{\alpha\times 2M}\int
d[\hat{Q}]\exp\left\{-\frac{N}{2}\rm Tr\hat{Q}^2+N \rm
Tr\ln\left( i\hat{Q}+
i \hat{\varepsilon}+\mu \openone \right)\right\},
\label{Q}
\end{equation}
} where we introduced the notation $\openone = \delta_{ij}\delta{mn}$.
Following Guhr~\cite{Guhr} we make the following change of variables in
Eq. (\ref{Q}): $\hat{Q} = \hat{Q}'-\hat{\varepsilon}+i \mu \openone$.  As
a result we obtain the following expression \wide{m}{
\begin{equation}
\langle Z^{\alpha}_0 (\mu)\rangle=c_{\alpha\times 2M}\int
d[\hat{Q}']\exp\left\{-\frac{N}{2}\rm
Tr(\hat{Q}'-\hat{\varepsilon}+i\mu \openone)^2+N \rm
Tr\ln\left( i\hat{Q}'\right)\right\}.
\label{Q'}
\end{equation}
}

The matrix $\hat{Q}'$ can be diagonalized using a unitary matrix $\hat{U}$
\begin{eqnarray}
  \label{eq:ULambda}
\hat{Q}'&=&U^{-1}\hat{\Lambda}U, \\
  \hat{\Lambda}&=&
\delta_{ij}\delta_{nm}\lambda_n^j. \nonumber
\end{eqnarray}

In terms of the new variables $\hat{U}$ and $\hat{\Lambda}$ the
integration measure becomes
\begin{equation}
  \label{eq:jacobian}
d[\hat{Q}']=\Delta^2(\hat{\Lambda})d[\hat{\Lambda}]d[\hat{U}],
\end{equation}
where $d[\hat{\Lambda}]=\prod\limits_{i,n} d\lambda^i_n$ and $d[\hat{U}]$
denotes the invariant integration measure in the unitary group.  The
Jacobian of the transformation (\ref{eq:ULambda}) is expressed through the
Vandermonde determinant $\Delta(\hat{\Lambda})\equiv \prod '
(\lambda^i_m-\lambda^j_n)$, where the prime indicates that the product is
taken over non-coinciding pairs of indices ${\{i,m\} \neq \{j,n\}}$. In
the variables $\hat{\Lambda}$ and $d[\hat{U}]$ Eq.~(\ref{Q'}) acquires the
form \wide{m}{
\begin{equation}
\langle Z^{\alpha}_0 (\mu)\rangle=c_{\alpha\times 2M}\int
\Delta^2(\hat{\Lambda})
d[\hat{\Lambda}]d[\hat{U}]\exp\left\{-\frac{N}{2}\rm
Tr(\hat{U}^{-1}\hat{\Lambda}\hat{U}-\hat{\varepsilon}+i \mu
\openone)^2+N \rm
Tr\ln\left(i \hat{\Lambda} \right)\right\}
\label{L}
\end{equation}
}

The integration over the unitary group $d[\hat{U}]$ in Eq.~(\ref{L}) can
be carried out exactly using the integral first computed by
Harish-Chandra~\cite{Harish-Chandra57} and known in the physics community
as the Itzykson-Zuber integral.~\cite{Zuber} As a result we obtain
\wide{m}{
\begin{equation}
\langle Z^{\alpha}_0 (\mu)\rangle=c_{\alpha\times
2M}\left(\frac{\pi}{N}\right)^{\alpha M(\alpha2M-1)}\int
d[\hat{\Lambda}]
\left(\frac{\Delta(\hat{\Lambda})}{\Delta(\hat{\varepsilon})}\right)
\exp\left\{-N \rm Tr
\left(-\ln \left( i \hat{\Lambda}\right) +\frac{1}{2}
(\hat{\Lambda}-\hat{\varepsilon}+i \mu \openone )^2\right)\right\}.
\label{ZQ}
\end{equation}
} The Vandermonde determinant $\Delta(\hat{\varepsilon})=\prod '
(\varepsilon_n-\varepsilon_m)$ vanishes if the number of replicas $\alpha$
is greater than unity because of the presence of identical Matsubara
frequencies in different replicas.  Therefore the expression in
Eq.~(\ref{ZQ}) appears to diverge. This is not so however\cite{Mezard},
since the integral over $d[\hat{\Lambda}]$ also vanishes when eigenvalues
of the matrix $\hat{\varepsilon}$ become degenerate.  To regularize the
integral in Eq.~(\ref{ZQ}) we introduce a {\em diagonal} regulator matrix
$\hat{V}=\delta_{ij}\delta_{mn}V_i$ with $V_i \neq V_j$, and calculate the
expression \wide{m}{
\begin{equation}
  \label{eq:ZQreg}
\langle Z^{\alpha}_0(\hat{V}) \rangle =  c_{\alpha\times
2M}\left(\frac{\pi}{N}\right)^{\alpha M(\alpha2M-1)}\int
d[\hat{\Lambda}]
\left(\frac{\Delta(\hat{\Lambda})}{\Delta(\hat{\varepsilon}+\hat{V})}\right)
\exp\left\{-N \rm \sum\limits_{j,m}F(\lambda^j_m, x_{m,j})
\right\},
\end{equation}
} where the function $F(\lambda^j_m, x_{m,j})$ is defined by the equation
\begin{mathletters}
\label{eq:Fab}
\begin{eqnarray}
\label{F}
 F(\lambda^j_m, x_{m,j})&=&
-\ln( i \lambda^j_m) +\frac{1}{2}
(\lambda^j_m-x_{m,j})^2,  \\
x_{m,j}&\equiv& \varepsilon_m-i\mu+V_j.
\label{x}
\end{eqnarray}
\end{mathletters}
The ratio of the two Vandermonde determinants in Eq. (\ref{eq:ZQreg}) is
rendered finite due to the presence of the regulators $V_i$ in
Eq.~(\ref{x}):
\begin{equation}
\frac{\Delta(\hat{\Lambda})}{\Delta(\hat{\varepsilon}+\hat{V})}=
\prod\limits_{\{i,m\} \neq \{j,n\}} \frac{\lambda^i_m -
\lambda^j_n}{x_{m,i}-x_{n,j}}.
\label{Vandermonde}
\end{equation}
We therefore carry out the integration over $d[\hat{\Lambda}]$ in
Eq.~(\ref{eq:ZQreg}) at finite $\hat{V}$ and take the limit $\hat{V}\to 0$
at the end of the calculations.

For $N\gg 1$ the integration over $d[\hat{\Lambda}]$ in the r.h.s. of Eq.
(\ref{eq:ZQreg}) can be performed by the saddle point
method.~\cite{Guhr,ben} To find the saddle point value of $\lambda^i_{m}$
at $N \gg 1$ we may ignore the ratio of the Vandermonde determinants in
the pre-exponential factor in the right hand side of Eq.~(\ref{eq:ZQreg}).
This can be done even after we let the number of Matsubara frequencies
$2M$ go to infinity, as we shall do eventually.  We show in Appendix
\ref{temp} that the shift of the saddle point equation which arises from
the pre-exponential factor is small as $\delta/T$ and may be neglected.
We denote the saddle point value of $\lambda^i_m$ by $\bar{\lambda}^i_m$.
Minimization of the right hand side of Eq.~(\ref{F}) leads to the
following saddle point equation
\begin{equation}
  \label{eq:speq}
  \bar{\lambda}^i_m \left( \bar{\lambda}^i_m -x_{m,i}
  \right) =1 .
\end{equation}
This equation has two solutions for each $\bar{\lambda}^i_m$.
\begin{equation}
\bar{\lambda}^i_m =\lambda_\pm (x_{m,i})= \frac{x_{m,i}}{2}\pm
\sqrt{1+\frac{x_{m,i}^2}{4}}
\rm .
\label{eq:lambdapm}
\end{equation}
Therefore there are $2^{2M \alpha}$ saddle points of the integrand in
Eq.~(\ref{eq:ZQreg}).

It is clear from Eq.~(\ref{eq:ZQreg}) that the fluctuations of
$\lambda^i_m$ about its saddle point value, Eq.~(\ref{eq:lambdapm}) are of
order $1/\sqrt{N}$. Let us assume for a moment that both the temperature
$T$ and $V_i$ satisfy the condition $1/\sqrt{N}\ll T, V_i -V_j \ll 1$.
Then the ratio of the Vandermonde determinants in the pre-exponential
factor in Eq.~(\ref{eq:ZQreg}) can be replaced by its value at the saddle
point
\begin{equation}
\left.
  \frac{\Delta(\hat{\Lambda})}{\Delta(\hat{\varepsilon}+\hat{V})}\right|_{\rm
 sp}  =
\prod\limits_{\{i,m\} \neq \{j,n\}} \frac{\bar\lambda (x_{m,i}) -
\bar \lambda (x_{n,j})}{x_{m,i}-x_{n,j}}.
\label{Vandermondesp}
\end{equation}
Next, we formally extend the product in this equation to include the
diagonal terms $\{i,m\} = \{j,n\}$ by replacing the ratio in the r.h.s.
of Eq.~(\ref{Vandermondesp}) by $d\bar \lambda(x_{m,i})/d x_{m,i}$ to
obtain \wide{m}{
\begin{equation}
\left.
  \frac{\Delta(\hat{\Lambda)}}{\Delta(\hat{\varepsilon}+\hat{V})}\right|_{\rm  
  sp} =
\exp\left\{\frac{1}{2}\sum\limits_{im,nj}
\ln\left(\frac{\bar \lambda (x_{m,i})-
\bar \lambda(x_{n,j})}{x_{m,i}-x_{n,j}}
\right)\right\}
\prod\limits_{i,m}\left(\frac{d\bar \lambda(x_{m,i}
    )}{d\varepsilon_m}\right)^{-1/2} ,
\label{Vandermonde5}
\end{equation}
} where the sum over the replica and the Matsubara indices is unrestricted
and includes the terms with $\{i,m\}=\{j,n\}$.  The factor $1/2$ in the
exponent in Eq.~(\ref{Vandermonde5}) appears due to double counting of
terms in the summation.

We now substitute Eq.~(\ref{Vandermonde5}) into Eq.~(\ref{eq:ZQreg}) and
expand the function $F(\lambda^i_m, \varepsilon_m, V_i)$ in Eq.~(\ref{F})
to second order in the fluctuations around the saddle point
Eq.~(\ref{eq:lambdapm}). It is easy to show that
\begin{equation}
\left.\frac{\partial^2 F(\lambda^i_m, x_{m,i})}
{\partial{\lambda^i_m}^2}\right|_{\bar \lambda}
=\left( \frac{\partial\bar \lambda(x_{m,i})}{\partial
    \varepsilon_m}\right)^{-1}.
\end{equation}
Therefore the Gaussian integration over the fluctuations around the saddle
point, Eq.~(\ref{eq:lambdapm}) results in the product
$\prod\limits_{i.m}[(2\pi/N)
d\bar{\lambda}(x_{m,i})/d\varepsilon_m]^{1/2}$ which partially cancels the
last product in Eq.~(\ref{Vandermonde5}), and we obtain \wide{m}{
\begin{eqnarray}
\langle Z^{\alpha}_0 (\hat{V})\rangle &=& \sum\limits_{\{ \bar \lambda \}}
\exp\left\{-N \sum\limits_{m,i}F(\bar \lambda (x_{m,i}),
    x_{m,i}) +
\frac{1}{2}\sum\limits_{i,j;n,m}
\ln\left(\frac{\bar \lambda(x_{m,i})-
\bar \lambda(x_{n,j})}{x_{m,i}-x_{n,j}}
\right)\right\},
\label{QZ2}
\end{eqnarray}
} where $\sum\limits_{\{ \bar \lambda \}}$ denotes the sum over all the
saddle points given by Eq.~(\ref{eq:lambdapm}).

Although, to justify the saddle point procedure and to arrive at
Eq.~(\ref{QZ2}) we assumed $V_i-V_j \gg 1/\sqrt{N}$, the saddle point
result (\ref{QZ2}) is exact for the unitary ensemble and holds the way to
$V_i\to 0$. It was shown by Zirnbauer~\cite{Zirnbauer99} that this special
feature of the unitary ensemble is a consequence of the
Duistermaat-Heckman theorem.~\cite{Duistermaat82}

Equation (\ref{QZ2}) is free of divergences in the $\hat{V}\to 0$ limit.
For the replica-symmetric saddle points it is obvious because in the
``dangerous'' terms with $m=n$ the argument of the logarithm in
Eq.~(\ref{QZ2}) remain finite and equal to $\partial\bar
\lambda(\varepsilon_m-i\mu)/\partial \varepsilon_m$ in the $V_i\to 0$
limit. For the saddle points with broken replica symmetry it is necessary
to sum the contributions of all replica permutations of a given saddle
point in order to obtain a finite result.

Next, we take the number $2M$ of the Matsubara frequencies to infinity.
The resulting infinite sums over the Matsubara frequencies in
Eq.~(\ref{QZ2}) diverge and need to be regularized.  The method of
regularization can be inferred from the observation that the sums over the
Matsubara frequencies should be regarded as traces of operators in the
Matsubara space.  Let us consider the one particle Matsubara Green
function as an example
\begin{equation}
G_{x,y}(\tau_1 - \tau_2)=-\langle
\psi_{x}(\tau_1)\bar{\psi}_{y}
(\tau_2)\rangle,
\end{equation}
where the average is taken over the functional integral as in
Eq.~(\ref{Z}).  It is important to remember that the Fermion operators
${\hat \psi}$ and $\hat {\bar \psi}$ appearing in the Hamiltonian which
enter the operator expression for the partition function $Z_0(\mu,T)$ are
normal ordered.  Therefore, in the time-discretized version of the
functional integral for the partition function, Eq.~(\ref{Z}) the
Hamiltonian $\hat{H}_0$ is coupled to fermionic fields at different
moments of time: $\bar{\psi}(t_{i+1})\hat{H}_0 \psi(t_i)$. This forces us
to understand the traces of the Greens function as
\begin{equation}
\label{tracetime}
\rm Tr \hat{G}
=\int\limits^{\beta}_0 d\tau
\lim_{\eta \to +0} \sum\limits_{x=1}^{N} G_{x,x}
( -\eta),
\end{equation}
which coincides with the total number of particles in the system.  In the
frequency representation the regularized trace of the Green function is
given by
\begin{equation}
\label{tracefreq}
\rm Tr \hat{G}=T \lim_{\eta \to
+0}\sum\limits_{\varepsilon_n=-\infty}^{+\infty}\sum\limits_{x=1}^{N}
G_{x,x}(\varepsilon_n)e^{i\varepsilon_n\eta}.
\end{equation}
The exponential factor $e^{i\varepsilon_n\eta}$ in Eq.~(\ref{tracefreq})
is essential for the convergence of the sum over the Matsubara
frequencies.  Since the Hamiltonian $\hat{H}_0$ in Eq.~(\ref{Z}), over
which the averaging is performed, is normal ordered the traces of {\em
  all} the ensuing operators should be regularized by the introduction of
the factors $e^{i\varepsilon_n\eta}$ as in Eq.~(\ref{tracefreq}).

Thus, Eq.~(\ref{QZ2}), where the summations over Matsubara frequencies
should be understood according to the rule $\sum_{n} f(n) \to
\sum_{n=-\infty}^{\infty} f(n) \exp(i\eta \varepsilon_n)$, represents an
exact expression for the averaged replicated partition function for $N \to
\infty$.

Below we will evaluate Eq.~(\ref{QZ2}) at $T \gg \delta$. In this case the
sum in Eq.~(\ref{QZ2}) is dominated by the saddle point which corresponds
to the lowest value of the exponent in Eq.~(\ref{QZ2}). The contributions
of all the other saddle points are exponentially small in $T/\delta$.  The
leading term in the sum in Eq.~(\ref{QZ2}) corresponds to the saddle point
\begin{equation}
\bar{\lambda}^i_m = \lambda^0(x_{m,i})=\frac{x_{m,i}}{2}+
\sqrt{1+\frac{x_{m,i}^2}{4}}
\rm .
\label{lambda}
\end{equation}
Here the function $\lambda^0(z)=z/2+\sqrt{1+z^2/4}$ is understood as an
analytic function of $z$ in the complex plane with a branch cut from $-2i$
to $2i$ as in Fig.~\ref{fig1}, in particular the square root in
Eq.~(\ref{lambda}) changes sign when the real part of $z$, $\Re z$,
crosses zero.

To obtain the expression for $\langle Z^{\alpha}_0(\mu,T)\rangle $ we take
the limit $V_i\to 0$ in Eq.~(\ref{QZ2}) and perform the summation over the
replica indices. As a result we obtain \wide{m}{
\begin{eqnarray}
\langle Z^{\alpha}_0(\mu)\rangle &=&
\exp\left\{\lim_{\eta \to + 0} \left[-N \alpha
  \sum\limits_{m=-\infty}^{\infty} F(\lambda^0(\varepsilon_m -i \mu),
  \varepsilon_m -i \mu )  e^{i\varepsilon_m \eta}
\right. \right. \nonumber \\
  &+& \left. \left.
\frac{\alpha^2}{2}\sum\limits_{m,n=-\infty}^{\infty}
e^{i(\varepsilon_m
  +\varepsilon_n)\eta}\ln\left(\frac{\lambda^0(\varepsilon_m-i\mu )-
\lambda^0(\varepsilon_n-i\mu )}{\varepsilon_m-\varepsilon_n}
\right)\right]\right\}.
\label{ZQinf}
\end{eqnarray}
}

The first term in the exponent in this equation is linear in the number of
replicas $\alpha$ and corresponds to the average grand canonical potential
$\langle\Omega_0(T,\mu)\rangle$ which is evaluated in section
\ref{sec:freen}.  The second term in the exponent of Eq.~(\ref{ZQinf}) is
quadratic in $\alpha$ and describes the mesoscopic fluctuations of
$\Omega_0(T,\mu)$ which are studied in section \ref{sec:fluct}.

\subsection{Average grand canonical potential}
\label{sec:freen}

In this section we evaluate the average grand canonical potential
$\langle\Omega_0(\mu)\rangle$.  It is easy to show using
Eqs.~(\ref{eq:cumulants}) and (\ref{ZQinf}) that it is given by
\begin{equation}
  \label{eq:omega}
  \langle\Omega_0(\mu)\rangle =
N T \lim_{\eta \to 0}
  \sum\limits_{m=-\infty}^{\infty}
  F(\lambda^0(\varepsilon_m -i \mu), \varepsilon_m -i \mu
    )e^{i\varepsilon_m \eta} .
\end{equation}

\begin{figure}
\epsfysize =3cm
\centerline{\epsfbox{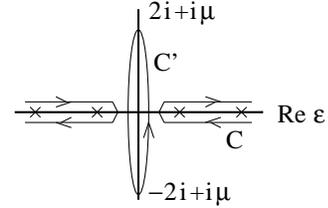}}
\caption{The initial sum over the Matsubara frequencies in
  Eq. (\ref{eq:omega}) is rewritten through the integral over the contour
  $C$ in Eq.~(\ref{contour}). Using the analytic properties of function
  $\lambda^0$ in Eq. (\ref{lambda}) the integration contour $C$ can be
  deformed to the  contour $C'$. }
\label{fig1}
\end{figure}

The function $F(\lambda^0(\varepsilon_m -i \mu), \varepsilon_m -i\mu)$ is
defined through Eqs. (\ref{F}) and (\ref{lambda}). It has a branch cut
from $-2i+i\mu$ to $2i+i\mu$ in the complex plane of the variable
$\varepsilon_m$ and is analytic everywhere else. It is convenient to
rewrite this sum in terms of the contour integral
\wide{m}{
\begin{equation}
\langle\Omega_0(\mu)\rangle
 = \frac{N}{4\pi i }\lim_{\eta \to 0} \oint\limits_C d\varepsilon
F(\lambda^0(\varepsilon_m -i \mu), \varepsilon_m -i \mu
  )
e^{i\varepsilon\eta}\left(\tan\left(\frac{\varepsilon}{2T}
\right)+i\right),
\label{contour}
\end{equation}
}
where the contour $C$ is shown in Fig. \ref{fig1}. Taking into account the
analytic properties of $F(\lambda^0(\varepsilon_m -i \mu), \varepsilon_m
-i \mu)$ we can deform the integration contour to $C'$ and integrate once
by parts to obtain
\wide{m}{
\begin{equation}
\label{C'}
\langle\Omega_0(\mu)\rangle =
\frac{N T}{2\pi i}\oint\limits_{C'}
d\varepsilon \left[ \frac{\varepsilon-i\mu}{2}
  -\sqrt{1+\frac{(\varepsilon-i\mu)^2}{4}}\right]
\ln\left(\frac{1+e^{-i\varepsilon/T}}{2}\right).
\end{equation}
}
Note that the  two terms in the
square brackets coincide with the trace (in the Hilbert space) of the
averaged single particle Green function. The latter is equal to the
retarded Green function on the right side of the branch cut and to the
advanced Green function on its left side. Therefore, upon the change of
variables $\varepsilon=-i(E-\mu)$ we can express the integral in
Eq.~(\ref{C'}) through the average single particle density of states
defined in Eq.~(\ref{dos})
\begin{equation}
\langle\Omega_0(\mu)\rangle
=-T\int\limits_{-2}^2dE\nu(E)\ln\left(\frac{1+e^{-(E-\mu)/T}}{2}\right) .
\label{Omega}
\end{equation}
This result is precisely what one expects for the average grand canonical
potential of non-interacting electrons.

\subsection{Mesoscopic fluctuations of the grand canonical potential}
\label{sec:fluct}

\begin{figure}
  \epsfysize =4cm \centerline{\epsfbox{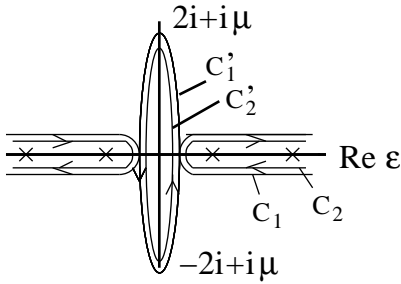}}
\caption{The initial sums over the Matsubara frequencies in
  Eq.~(\ref{eq:secondmoment}) are rewritten as the contour integrals in
  Eq.~(\ref{contour3}) over the contours $C_1$ and $C_2$ which are later
  deformed to the contours $C_1'$ and $C_2'$ in Eq.~(\ref{contour5}). }
\label{meso}
\end{figure}

In this section we consider the mesoscopic fluctuations of $\Omega_0(\mu)$
at $T\gg \delta$. From Eqs.~(\ref{eq:cumulants}) and (\ref{ZQinf}) we
immediately find that the second cumulant $\langle \langle \Omega_0(\mu)
\rangle \rangle$ is determined by the second term in the exponent of
Eq.~(\ref{ZQinf}) \wide{m}{
\begin{equation}
  \label{eq:secondmoment}
  \langle  \langle \Omega_0^2(\mu) \rangle \rangle =T^2
  \sum\limits_{m,n=-\infty}^{\infty} e^{i(\varepsilon_m
  +\varepsilon_n)\eta}\ln\left(\frac{\lambda^0(\varepsilon_m-i\mu )-
\lambda^0(\varepsilon_n-i\mu )}{\varepsilon_m-\varepsilon_n}
\right)
\end{equation}
} As in the previous section it is convenient to rewrite the summation
over the Matsubara frequencies in the last formula in terms of the contour
integral \wide{m}{
\begin{equation}
\langle  \langle \Omega_0^2(\mu)\rangle \rangle =\frac{1}{(4\pi i
)^2}\oint \limits_{C_1} \oint \limits_{C_2} d\varepsilon_1 d\varepsilon_2
\ln\left(\frac{\lambda^0(\varepsilon_m-i\mu )-
\lambda^0(\varepsilon_n-i\mu )}{\varepsilon_m-\varepsilon_n}
\right)\left(\tan\frac{\varepsilon_1}{2T}
+i\right)\left(\tan\frac{\varepsilon_2}{2T}
+i\right)e^{i(\varepsilon_1 + \varepsilon_2)\eta}
\label{contour3}
\end{equation}
} where the contours of integration $C_1$ and $C_2$ are shown in Fig.
\ref{meso}.
In Eq.~(\ref{contour3}) the function $\lambda^0(z-i\mu)$ was defined in
Eq.~(\ref{lambda}) and is understood as a function of $z$ which is
analytic everywhere in the complex plane except for a branch cut from
$-2i+i\mu$ to $2i+i\mu$ as in Fig.~\ref{meso}.  We therefore can deform
the integration contours $C_1$ and $C_2$ to $C_1^{'}$, $C_2^{'}$.  After
that we integrate by parts once with respect to both $\varepsilon_1$ and
$\varepsilon_2$ and obtain \wide{m}{
\begin{equation}
\langle  \langle \Omega_0^2(\mu)\rangle \rangle=
-\frac{ T^2}{(2 \pi )^2}\oint\limits_{C_1^{'}}
\oint\limits_{C_2^{'}}d\varepsilon_1d\varepsilon_2
G_c(i\varepsilon_1 +\mu,i\varepsilon_2+\mu)
\ln\left(\frac{1+e^{-i\varepsilon_1/T}}{2}\right)\ln\left(
\frac{1+e^{-i\varepsilon_2/T}}{2}\right),
\label{contour5}
\end{equation}
} where we introduced the function \wide{m}{
\begin{equation}
G_c(i\varepsilon_1+\mu,i\varepsilon_2+\mu)=\frac{\partial^2
  }{\partial\varepsilon_1  \partial \varepsilon_2}\ln\left(
  \frac{\frac{\varepsilon_1-i\mu}{2}+
    \sqrt{1+\frac{(\varepsilon_1-i\mu)^2}{4}}-
    \frac{\varepsilon_2-i\mu}{2}-\sqrt{1+\frac{(\varepsilon_2
        -i\mu)^2}{4}}}  {\varepsilon_1-\varepsilon_2}\right).
\label{contour4}
\end{equation}
} The function $G_c(\varepsilon_1,\varepsilon_2)$ is equal to the smooth
part (i.e. not containing the oscillations on the scale of the mean level
spacing $\delta$) of the connected two point correlator of Green functions
at energies $i \varepsilon_1+\mu$ and $i \varepsilon_2+\mu$ which was
obtained by Brezin and Zee.~\cite{Zee}

Using the fact that the square root in $\lambda_0(\varepsilon_i-i\mu)$ has
a different sign on the different sides of the contour $C_i^{'}$ we can
express $ \langle \langle \Omega_0(\mu)^2 \rangle \rangle$ through the
integral over, say the right side of the contours $C_1^{'}$ and $C_2^{'}$
only.  Introducing the variables $E_1=i\varepsilon_1+\mu$ and
$E_2=i\varepsilon_2+\mu$ we obtain \wide{m}{
\begin{equation}
\langle  \langle \Omega_0^2(\mu)\rangle \rangle=T^2
\int\limits_{-2}^2
\int\limits_{-2}^2dE_1dE_2
\tilde \rho_c(E_1,E_2)
\ln\left(\frac{1+e^{-(E_1-\mu)/T}}{2}\right)\ln\left(
\frac{1+e^{-(E_2-\mu)/T}}{2}\right),
\label{contour6}
\end{equation}
} where $\tilde \rho_c(E_1,E_2)$ denotes the smooth part of the connected
density of states correlator~\cite{Zee}
\begin{equation}
\tilde \rho_c(E_1,E_2)=\frac{E_1E_2-4}
{2\pi^2 (E_1-E_2)^2\sqrt{(4-E_1^2)(4-E_2^2)}}.
\end{equation}
This expression has a singularity when $E_1-E_2\to 0$ which leads to the
divergence of the integral in Eq.~(\ref{contour6}).  If we replace $\tilde
\rho_c(E_1,E_2)$ in Eq.~(\ref{contour6}) by the exact connected density
correlation function $\rho_c(E_1,E_2)$ which does not diverge at
$E_1-E_2\to 0$ the integral in Eq.~(\ref{contour6}) becomes finite and
gives the exact expression for the second cumulant of the grand canonical
potential.  Since we have retained only one term corresponding to the
lowest energy saddle point $\{\lambda_0\}$ in the sum over the saddle
points in Eq.~(\ref{QZ2}), the expression in Eq.~(\ref{contour6}) is
missing the oscillatory part of the two point level density correlator.
These oscillatory terms arise from the other saddle points in
Eq.~(\ref{QZ2}) and cut off the divergence of the integral at the scale of
the mean level spacing $E_1-E_2 \approx \delta \sim 1/N$. Therefore the
integral in Eq.~(\ref{contour6}) should be cut off at $E_1-E_2 \sim 1/N$
which leads to the variance $\langle \langle \Omega_0^2(\mu)\rangle
\rangle \sim N^2 \ln N$. This result is to be expected since the main
contribution to the thermodynamic potential arises from levels deep below
the Fermi surface.  The fluctuation of the number of such levels is
proportional to $\ln N$ and their typical energy is of order $-N$.

\section{Coulomb interaction and tunneling}
\label{interaction}

In this section we consider a disordered metallic grain coupled by a
tunneling contact to a clean metallic lead and capacitively coupled to
a metallic gate as in Fig.~\ref{dot}.  The Thouless energy of the
grain is assumed to be greater than the charging energy, $E_T >> E_C$.
In addition, we assume that the dot is placed in a magnetic field,
such that the cooperon gap exceeds the charging energy $DeH/(\hbar c)
\gg E_C$.  It was shown by Efetov~\cite{Efetov97} that under these
conditions the ensemble of disorder potentials in the grain can be
described by the unitary random matrix ensemble.  At the same time we
assume that the Zeeman splitting of electron energy levels, $\hbar
eH/mc$ is smaller than the temperature $T$ and may be neglected.  We
therefore assume that each orbital level is doubly degenerate.

The single particle levels in the grain are broadened due to tunneling
into the lead.  We assume that the tunneling contact is broad, so that
each single particle state in the dot is coupled to many lead states.
 In this case the probability distribution $P(\Gamma_x)$ of level
half-widths $\Gamma_x$ is sharply peaked about the mean value
$\Gamma$.  We can therefore neglect the fluctuations of $\Gamma_x$
and consider them to be equal to the mean value $\Gamma$ which can be
expressed through the dimensionless conductance of the contact $g_T$
as
\begin{equation}
\Gamma=\frac{g_T\delta}{8\pi}.
\label{Gamma1}
\end{equation}

We assume that the temperature satisfies the conditions $\delta \ll T \ll
E_C$  and focus on the Coulomb blockade oscillations in the
thermodynamic quantities of the grain at $g_T \gg 1$ and at a finite
mean level spacing $\delta$. The quantity of particular interest is
the differential capacitance of the dot $C(q)$ which can be expressed
through the grand canonical potential $\Omega(q)$ of the dot at gate
voltage $q$ as
\begin{equation}
  \label{eq:capacitdef}
  C(q) = \kappa \frac{\partial^2 \Omega(q)}{ \partial q^2},
\end{equation}
where $\kappa$ is a factor depending on the geometry of the dot.  In
addition to the average differential capacitance and the grand
canonical potential, Eq.~(\ref{eq:cumulants}), we study correlations
of thermodynamic quantities at different values of the gate voltage,
$q$ and $q'$. For this purpose we calculate the product of replicated
partition functions at different values of the gate voltage $\langle
Z^\alpha(q) Z^{\alpha'}(q')\rangle$ averaged over the ensemble of
random Hamiltonians with the distribution (\ref{distribution}).
The correlators $ \langle \Omega^n (q) \Omega^m(q') \rangle $ can then
be obtained from the formula \wide{m}{
\begin{equation}
  \label{eq:connectcorr}
  \langle \Omega^n (q) \Omega^m(q') \rangle
 = \left. (-T)^{n+m} \left(\frac{\partial}{\partial \alpha} \right)^n
\left(\frac{\partial}{\partial \alpha'} \right)^m
   \langle Z^\alpha(q)  Z^{\alpha'}(q')\rangle
\right|_{\alpha,\alpha'\to 0}.
\end{equation}
}

\subsection{Average thermodynamic potential}
\label{sec:coulaverage}

In the absence of the Coulomb interaction but at a finite tunneling rate
$\Gamma$ we can write the partition function $Z_\Gamma^0(\mu,T)$, where
$\mu$ is the chemical potential of electrons in the lead, as a functional
integral over the fermionic fields $\psi$ \wide{m}{
\begin{equation}
Z_\Gamma^0(\mu,T)=\int\prod\limits_{x=1}^{N}\prod\limits_{n=-\infty}^{\infty}
d\bar{\psi}^\sigma_{x,n}d{\psi^\sigma_{x,n}}
\exp\left[\sum\limits_{n,m=-\infty}^{\infty}\sum\limits_{x,y=1}^N
  \sum\limits_{\sigma}
\bar{\psi}^\sigma_{x,n}\left(i \hat{\varepsilon}+i \Gamma\rm sgn(\hat
  \varepsilon) - \delta_{nm}(\hat{H}_{0,xy}-\mu \delta_{xy})
\right)\psi^\sigma_{y,m} \right].
\label{Zg}
\end{equation}
} Here the index $\sigma$ denotes the spin of the particle.

In the presence of the Coulomb interaction, Eq.~(\ref{eq:cint}), the
replicated partition function $Z^\alpha(q)$ may be written as a functional
integral over the fermionic fields similar to that in Eq.~(\ref{Zg}).
However, in this case the exponent acquires a quartic term in the fermion
fields $\psi$.  A convenient way to proceed is to decouple the interaction
via the Hubbard-Stratonovich transformation by introducing an auxiliary
field $V(\tau)$.  Following Finkelstein~\cite{Fin} we introduce such
auxiliary fields $V_j$ in each replica, \wide{m}{
\begin{equation}
\exp\left\{-\sum\limits_{i=1}^{\alpha}
\int\limits_0^{\beta}E_C(N_j(\tau)-q)^2d
\tau\right\}=
\int d[V]\exp\left\{-\sum\limits_{j=1}^{\alpha}\left(
\int\limits_0^{\beta}d\tau
\frac{V^2_j(\tau)}{4E_C}-i\int
\limits_{0}^{\beta}V_j(\tau)(N_j(\tau)-q)d\tau\right)\right\},
\label{decouple}
\end{equation}
} where $N_j(\tau) \equiv \sum_{x,\sigma}
\bar{\psi}^\sigma_{x,j}(\tau)\psi^\sigma_{x,j}(\tau)$.  Next we denote the
static part of $V_j(\tau)$ by $V_j$ and write the integral over $V_j$ as a
sum of integrals labeled by a set of winding numbers $\{W\}$~\cite{AES}
\[ \prod\limits_{j} \int\limits_{-\infty}^{\infty} d V_j  =
\sum\limits_{\{W\}}\prod\limits_{j} \int\limits_{\pi T(2 W_j-1)}^{\pi T(2
  W_j+1)} d V_j .
\]
For each set of winding numbers we express $V_j(\tau)$ as
\begin{equation}
  \label{eq:phidef}
  V_j(\tau)=V_j -2\pi TW_j +\dot{\phi}_j(\tau),
\end{equation}
where the phases $\phi_j(\tau)$ satisfy the periodicity condition
$\phi_j(\beta)=\phi_j(0)+2\pi W_j$.  Then, making the gauge transformation
$\psi(\tau) \to \psi(\tau)\exp(i\phi(\tau))$, using Eqs.~(\ref{decouple}),
(\ref{Zg}) and averaging over the random matrix $\hat{H}_0$ with the
distribution function defined in Eq.~(\ref{distribution}) we write the
ensemble averaged replicated partition function $\langle
Z^{\alpha}(q)\rangle$ as a sum over the winding numbers \wide{m}{
\begin{equation}
\langle Z^{\alpha}(q)\rangle=\sum\limits_{\{W\}}
\prod\limits_{j=1}^{\alpha}\int\limits_{\pi T(2 W_j-1)}^{\pi
  T(2 W_j+1)}
\frac{dV_{j} }{\sqrt{4\pi E_c T}}
e^{-\frac{V^2_{j}}{4TE_c}
-iq\frac{V_{j}}{T}}
\int
 D\hat{Q}d[\phi] e^{-\frac{N}{2}{\rm Tr}\hat{Q}^2+ N {\rm
Tr}\ln\left( i\hat{Q}+ i\hat{J}\right) -
\int\limits_{0}^{\beta}\frac{(\dot{\phi}_j-2\pi T W_j)^2d\tau}{4E_C}} .
\label{Qint}
\end{equation}
} Here the trace is taken over the spin, the replica and the Matsubara
space.  The $\hat Q$-matrix was introduced to decouple the quartic
interaction of fermions arising from the averaging over $\hat H _0$,
$D\hat{Q}$ denotes the normalized integration measure, and we introduced
the operator
\[
\hat{J}=\hat{\varepsilon}-i\mu + \left[V_j - 2\pi T W_j+ \Gamma
  e^{-i\phi_j(\tau)}\Lambda_{\tau,\tau'}e^{i\phi_j(\tau')}\right]\delta_{ij},
\]
where $\hat{\varepsilon}=i\partial_{\tau}$, and $\Lambda_{\tau,\tau'}=
-i/\sin[\pi T(\tau-\tau')]$.

To perform the integration over $\hat Q$ in Eq.~(\ref{Qint}) we proceed as
in section \ref{noninteract}: we make a change of variables
$\hat{Q}\rightarrow \hat{Q}'-\hat{J}$ and parameterize the matrix
$\hat{Q}'$ in terms of the unitary matrix $\hat U$ and the diagonal matrix
$\hat{\Lambda} $ as $\hat{Q}'=\hat{U^{-1}}\hat{\Lambda}\hat{U}$. The
ensuing integral over the unitary matrices $\hat U$ is of the
Itzykson-Zuber type and can be evaluated exactly. The result is expressed
through the eigenvalues $J_m^i$ of the operator $\hat{J}$.  The operator
$\hat{J}$ is not diagonal in the Matsubara space but may still be
diagonalized using a unitary transformation.  The index $m$ can be
identified with the Matsubara index only if $\dot{\phi}=2 \pi T W$.  We
then perform the integration over $\hat{\Lambda}$ by the saddle point
method which for the unitary ensemble gives an exact result for $N \gg 1$.
The saddle point equation for the eigenvalues $\lambda_{m,\sigma}^i$ is
given by $\bar \lambda_{m,\sigma}^i(\bar \lambda_{m,\sigma}^i-J_m^i)=1$
and has two solutions $\lambda_{\pm}(J_m^i)$ defined in
Eq.~(\ref{eq:lambdapm}).  For $T \gg \delta$ the integration over the
static components $V_j$ can be carried out in the saddle point
approximation.~\cite{KamenevGefen} The saddle point equation reads
\begin{equation}
  \label{eq:lambdacharge}
 - i N T \sum_{n,\sigma} (\lambda^j_{n,\sigma} - J^j_n)  e^{i\eta
\varepsilon_n}= q - \frac{iV_j}{2E_C}.
\end{equation}
The left hand side of Eq.~(\ref{eq:lambdacharge}) may be interpreted as
the number of particles in the dot with the chemical potential $\mu +
i(V_j - 2\pi T W_j)$.  Note that in order to have a reasonably small
charging energy the imaginary part of the expression in the r.h.s. of this
equation should be at most of order unity.  Since the prefactor in the
l.h.s. $N T=\pi T/\delta$ is much greater than unity according to our
assumptions, the saddle point must have an equal number of $\lambda_+$ and
$\lambda_-$ in order to satisfy (\ref{eq:lambdacharge}).  Such a saddle
point with the lowest free energy is given by
\begin{equation}
\lambda_0(J_m^i) =\frac{J_{m}^i}{2} + {\rm sgn }\, \varepsilon_m \;
\sqrt{1+\frac{(J_m^i)^2}{4}} .
\label{lambda5}
\end{equation}
All other saddle points satisfying Eq.~(\ref{eq:lambdacharge}) are given
permutations between $\lambda_+$ and $\lambda_-$ in (\ref{lambda5}) such
that the numbers of $\lambda_+$ and of $\lambda_-$ are equal. These saddle
points provide only small contributions to the partition function $\sim
\exp(-T/\delta)$ and will be neglected.

We first solve Eq.~(\ref{eq:lambdacharge}) for $V_j$ at $\phi_j(\tau)=2\pi
W_j \tau$.  In this case the operator $\hat{J}$ is diagonal in the
Matsubara basis and its eigenvalues are given by
\[
J^j_m=\varepsilon_m -i\mu +V_j-2\pi T W_j +\Gamma {\rm sgn} \left(
  \varepsilon_m -2\pi T W_j\right).
\]
Substituting this expression and (\ref{lambda5}) into
Eq.~(\ref{eq:lambdacharge}) we find in the leading order in $\delta/E_C$
and $\delta/T$
\begin{equation}
V_j= 2\pi T W_j -i\frac{\delta}{2} \left[ q-N_0(\mu)\right],
\label{eq:V_j_sp}
\end{equation}
where $N_0(\mu)$ is the number of particles in the dot at the chemical
potential $\mu$ given by the l.h.s. of (\ref{eq:lambdacharge}) at
$V_j=W_j=0$.  To arrive at Eq.~(\ref{eq:V_j_sp}) we have used that
$\partial N_0(\mu)/\partial \mu = 2/\delta $. It is easy to convince
oneself that Eq.~(\ref{eq:V_j_sp}) holds for an arbitrary phase
configuration $\phi_j$ in the topological class $W_j$.  We therefore can
write the operator $\hat{J}$ at the saddle point (\ref{lambda5}) as
\begin{equation}
  \label{eq:J}
\hat{J}=\hat{\varepsilon}-i\frac{\delta}{2} \left[ q-N_0(\mu)\right] +
\Gamma e^{-i\phi_j(\tau)}\Lambda_{\tau,\tau'}e^{i\phi_j(\tau')}\delta_{ij}.
\end{equation}

Retaining only the leading saddle point (\ref{lambda5}) for each set of
winding numbers $\{ W\}$ and integrating over the gaussian fluctuations of
$V_j$ about the saddle point solutions (\ref{eq:V_j_sp}) and using the
fact that $ \int\limits_{0}^{\beta} (\dot{\phi}_j-2\pi T W_j)^2d\tau =
\int\limits_{0}^{\beta}\dot{\phi}_j^2d\tau + 4\pi^2 T W_j^2$ we can write
the replicated partition function as a sum over the winding numbers
\begin{eqnarray}
    \label{eq:windingsum}
          \langle Z^{\alpha}(q)\rangle &=& \left( \case{\delta}{4E_C}
        \right)^{\alpha/2}  \sum\limits_{\{W\}}
        \int d[\phi]
        Z^{\alpha}_{\{ W \}}\; e^{-
        \int\limits_{0}^{\beta}\frac{\dot{\phi}_j^2d\tau}{4E_C}},
\end{eqnarray}
where $Z^{\alpha}_{\{ W \}}$ and $\Delta_{\{W\}}[\phi]$ are given by the
expressions
\begin{eqnarray}
\label{eq:intsp}
Z^{\alpha}_{\{ W \}}&=&
\Delta_{\{W\}}[\phi]e^{- \sum_j \left[ 2\pi iqW_j +
 {\cal F}_{\{ W \}}(
 \phi_j )/T  \right]},  \\
\Delta_{\{W\}}[\phi]&=&\prod\limits_{i,j;n,m}\left(\case{
  \lambda_0(J_{m}^i) - \lambda_0(J_{n}^j)}{J_m^i-J_n^j}\right)^2 .
\label{eq:Vanddef}
\end{eqnarray}
Here we introduced the notation $ {\cal F}_{\{ W \}}( \phi_j )$ for the
free energy of the dot at the saddle point (\ref{lambda5})
\begin{equation}
  {\cal F}_{\{ W \}}( \phi_j )=NT
  \sum\limits_{n,\sigma} F(\lambda_0(J_{n}^j), J_n^j),
\label{free5'}
\end{equation}
where $F(\lambda_0(J_{n}^j), J_n^j) $ was defined in Eq.~(\ref{F}).  The
Vandermonde determinant in Eq.~(\ref{eq:Vanddef}) is generally finite.
This is so because the eigenvalues $J_m^i$ are non-degenerate due to the
presence of the auxiliary fields $\phi_i$.

For the winding number $W_j$ the minimum of the free energy ${\cal
  F}^j_{\{ W \}}( \phi )$ is achieved on the instanton phase configuration
$\phi_z(\tau)$ which can be expressed using complex variable $u=\exp(i
2\pi T \tau)$ as ~\cite{Nazarov},
\begin{equation}
\exp\left[i\phi_z(\tau)\right]=\prod\limits_{\rho =1
 }^{W_j}\frac{1-u^{-1}z_\rho}{1-uz_\rho^*}.
\label{eq:inst}
\end{equation}
Here the complex instanton parameters $z_\rho$ satisfy the inequality
$|z_\rho|<1$ for $W_j>0$, and $|z_\rho|>1$ for $W_j<0$.  The free energy
${\cal F}_{\{ W \}}( \phi_z )$ on the instanton configuration
(\ref{eq:inst}) is independent of $z_\rho$ and is given by
\begin{equation}
\label{minimum}
\frac{{\cal F}_{\{ W \}}( \phi_z )}{T}=
\frac{g_T}{2}|W_j|.
\end{equation}

It is clear from Eqs.~(\ref{minimum}), (\ref{eq:intsp}) that the largest
term in the sum over $\{W\}$ in Eq.~(\ref{eq:windingsum}) has a monotonic
dependence on the gate voltage, $q$ and corresponds to $\{W\}=\{0\}$ with
all $W_j=0$.  The leading oscillatory contributions to this sum arise from
the terms with $\{W\}=\{l\}$, $l=\pm 1$, having only one non-zero
$W_j=l=\pm 1$ and can be chosen in $\alpha$ ways by permutations between
the replicas.  Retaining only these terms we obtain \wide{m}{
\begin{equation}
\frac{\langle  Z^\alpha(q) \rangle}{\langle
  Z^\alpha(q) \rangle_{\{0\}} }=
1 +  \alpha\sum \limits_{l=\pm 1} e^{2\pi
    ilq} \frac{\int d[\phi]\Delta_{\{l\}}[\phi]\exp\left\{-
   \sum_{j}\left(
\frac{{\cal F}_{\{ l\}}( \phi_j )}{T}
+\int_0^{\beta}\frac{\dot{\phi}_j^2 d\tau}{4E_C}\right)\right\}}
{\int d[\phi]\Delta_{\{0\}}[\phi]\exp\left\{-\sum_{j}\left(
\frac{{\cal F}_{\{0\}}( \phi_j )}{T}
+ \int_0^{\beta}\frac{\dot{\phi}_j^2 d\tau}{4E_C}\right)\right\}},
\label{zz}
\end{equation}
} In Eq.~(\ref{zz}) we need to integrate over the fluctuations of $\phi$
around the instanton configuration Eq.~(\ref{eq:inst}). To this end we
write $\phi=\phi_z + \tilde \phi$, where $\tilde \phi$ represents the
massive modes of the dissipative action in Eq.~(\ref{zz}) The integration
over the zero modes of the dissipative action (\ref{zz}) should be
performed with the measure $d^2z/(1-|z|^2)$ for $|z|<1$, and
$d^2z/[|z|^2(|z|^2-1)]$ for $|z|>1$. All the other fluctuations $\tilde
\phi$ have a large mass of order $g_T$.  Therefore the integration over
them may be performed in the gaussian approximation.  Moreover, in
carrying out this integration one may replace the determinants
$\Delta[\phi]$ in Eq.~(\ref{zz}) by their values $\Delta[\phi_z]$ on the
instanton configurations (\ref{eq:inst}) due to their weak dependence on
$\tilde \phi$.  In this approximation the gaussian integrals over the
massive modes $\tilde \phi$ in different replicas in Eq.~(\ref{zz})
factorize.  For the replicas with vanishing winding numbers the integrals
in the numerator cancel with those in the denominator.  We are therefore
left with the expression involving only one replica in which the instanton
is present,
\begin{equation}
\frac{\langle  Z^\alpha(q) \rangle}{\langle
  Z^\alpha(q) \rangle_{\{0\}} }=
1 +  2 \alpha\Re  e^{2\pi i q}
  \int\limits_{|z|<1}\frac{d^2z f(z) }{(1-|z|^2)} 
  \frac{\Delta_{\{l\}}[\phi_z]}{\Delta_{\{0\}}[0]},
\label{eq:Zz}
\end{equation}
where $f(z)$ denotes the ratio of the integrals over the massive modes
with and without the instanton \wide{m}{
\begin{equation}
  \label{eq:fdef}
 f(z)\equiv \exp\left\{-\frac{2\pi^2
        T|z|^2}{E_C(1-|z|^2)}\right\}
 \frac{\int d[\tilde \phi]\exp\left\{-
   \frac{{\cal F}_{\{ 1\}}( \phi_z+\tilde\phi )}{T}
+\int_0^{\beta}\frac{(\dot{\tilde \phi})^2
  d\tau}{4E_C}\right\}} 
{\int d[\phi]\exp\left\{-
\frac{{\cal F}_{\{0\}}( \phi )}{T}
+ \int_0^{\beta}\frac{\dot{\phi}^2 d\tau}{4E_C}\right\}}.
\end{equation}
} To arrive at Eqs.~(\ref{eq:Zz}), (\ref{eq:fdef}) we used that
$\int_0^{\beta}\frac{\dot{\phi}_z ^2 d\tau}{4E_C}=\case{2\pi^2
  T|z|^2}{E_C(1-|z|^2)}$ and the fact that the terms with $l = \pm 1$ in
Eq.~(\ref{zz}) are complex conjugates of each other.

In order to evaluate the ratio of the Vandermonde determinants in
Eq.~(\ref{eq:Zz}) we need to find the eigenvalues of the operator
(\ref{eq:J}) on the instanton configurations.  This is done in Appendix
\ref{sec:eigenJ}. We then show in Appendix \ref{sec:Vandermondeinnst} that
in the limit $\alpha \to 0$ the ratio
$\Delta_{\{1\}}[\phi_z]/\Delta_{\{0\}}[0]$ is equal to unity.

The logarithmic divergence of the integral over $z$ for short instantons,
$|z| \to 1$, in Eq.~(\ref{eq:Zz}) is cut off at $1-|z|^2 \sim T/E_C$ by
the first term in $f(z)$, Eq.~(\ref{eq:fdef}).  The ratio of the integrals
over the massive modes in (\ref{eq:fdef}) is evaluated in Appendix
\ref{fluct}.  For very long instantons, $|z|\to 0$, the result is given by
$\frac{g_T^2E_C}{2\pi(\pi T + \Gamma)}$, whereas for short instantons,
$|z| \to 1$, and $\Gamma \gg T$ the result is $\frac{g_T^2E_C}{2\pi^2
  T}(1-\frac{\Gamma}{\pi T}\left(1-|z|^2\right) \ln(1+\Gamma/\pi T)) $.
To evaluate the integral (\ref{eq:Zz}) with logarithmic accuracy we may
interpolate $f(z)$ for the intermediate instanton lengths between the
two asymptotics as
\begin{equation}
  \label{eq:fresult}
f(z)=\frac{g_T^2E_C \exp\left\{-\frac{2\pi^2
        T|z|^2}{E_C(1-|z|^2)}\right\}}{2\pi^2 T(1+\case{\Gamma}{\pi T}\left[ 1-|z|^2\right ])}.
\end{equation}

Using this and integrating over $z$ in Eq.~(\ref{eq:Zz}) we obtain for the
oscillatory part of the {\em ensemble averaged} thermodynamic potential
\begin{equation}
\label{2}
 \langle\Omega(q)\rangle_{\rm osc}=
- \frac{g_T^2E_C}{\pi}\ln\left[\frac{E_C}{T +
    \Gamma}\right]\exp\left(-\frac{g_T}{2}\right) \cos(2\pi q). 
\end{equation}
Eq.~(\ref{2}) is the main result of this section.  It is applicable in the
temperature range $\delta \ll T \ll E_C$.  For $\Gamma \ll T$ it coincides
with the one instanton approximation of Ref.~\onlinecite{Grabert}.  The
result of Eq.~(\ref{2}) was obtained in the single instanton approximation
which holds for $T\gg g_T^2 E_C \exp(-\case{g_T}{2})$.

\subsection{Correlation function
  $\langle \Omega(q)\Omega(q')\rangle$}
\label{sec:intcorrelations}

In this section we consider the correlation function $\langle
\Omega(q)\Omega(q')\rangle$ of the thermodynamic potentials at different
values $q$ and $q'$ of the gate voltage. For this purpose we calculate the
average product of the replicated partition functions at two values of the
gate voltage, $\langle Z^\alpha (q) Z^{\alpha'}(q')\rangle $.  Repeating
the steps of section \ref{sec:coulaverage} we can write $\langle Z^\alpha
(q) Z^{\alpha'}(q')\rangle $ as a sum over the sets of winding numbers
$\{W\}$ and $\{W'\}$ \wide{m}{
\begin{equation}
  \label{eq:spcorr}
\langle
Z^\alpha (q) Z^{\alpha'}(q')\rangle= \sum\limits_{\{W\}}
\sum\limits_{\{W'\}}
\int d[\phi]
e^{-\sum\limits_{j=1}^{\alpha+\alpha'}
\int\limits_{0}^{\beta}\frac{\dot{\phi}_j^2d\tau}{4E_C}}
Z^{\alpha}_{\{ W\}} Z^{\alpha'}_{\{ W' \}}
    \prod\limits_{n,m}\prod\limits_{i=1}^\alpha
    \prod\limits_{j=1}^{\alpha'}
  \left(\frac{ \lambda_0(J_{m}^i)
- \lambda_0(J'^j_{n})}{J_m^i-J'^j_n}
\right)^4,
\end{equation}
} where $Z^{\alpha}_{\{W\}}$ and $Z^{\alpha'}_{\{ W \}}$ are given by
Eq.~(\ref{eq:intsp}).  The primed quantities in Eq.~(\ref{eq:spcorr})
refer to the set of replicas pertaining to the partition function at the
gate voltage $q'$.  If one replaces the product (Vandermonde determinant)
in Eq.~(\ref{eq:spcorr}) by unity this equation reproduces the product of
the averaged replicated partition functions. The deviations of this
product from unity describe correlations of the partition functions in the
different sets of replicas.

For a given set of winding numbers $\{W\}$ the functional integral over
the phases is dominated by configurations of $\phi$ close to the instanton
trajectories, Eq.~(\ref{eq:inst}). As in the previous section, when
integrating over the massive fluctuations $\tilde \phi$ about the
instanton configurations we may neglect the dependence of the Vandermonde
determinant in Eq.~(\ref{eq:spcorr}) on $\tilde \phi$ and evaluate it on
the instanton configuration.  The dominant contribution to the sum over
$\{W\}$ and $\{W'\}$ in Eq.~(\ref{eq:spcorr}) has a monotonic dependence
on the gate voltages $q$ and $q'$ and arises from the term with all
$W_j=0$.  The largest oscillatory contributions to this sum arise from the
terms with $\{W\}=\{l\}$ and $\{W'\}=\{l'\}$, $l,l'=\pm 1$, having only
one non-zero $W_j = \pm 1$ in each set of replicas.  Such terms can be
chosen in $\alpha \alpha'$ ways.  It can be shown, see Appendix
\ref{sec:Vandermondeinnst}, that the Vandermonde determinant in
Eq.~(\ref{eq:spcorr}) on the instanton configurations with $l=l'=\pm 1$
goes to unity in the limit $\alpha, \alpha' \to 0$. Such terms therefore
do not contribute to the irreducible correlator of the thermodynamic
potentials.  The contributions of the terms with $l=-l'=\pm 1$ are complex
conjugates of each other.  Using this property, denoting the Vandermonde
determinant in the $\alpha,\alpha' \to 0$ limit by $\Delta_{l,l'}$, and
using Eq.~(\ref{eq:connectcorr}) we can write the leading oscillatory term
in the irreducible correlator of thermodynamic potentials as \wide{m}{
\begin{equation}
\langle\langle\Omega(q)\Omega(q')\rangle\rangle_{\rm osc}
=2 \Re  e^{2\pi i (q-q')} 
  \int\limits_{|z'|,|z|<1}\frac{d^2z d^2z' f(z)
    f(z')}{(1-|z|^2)(1-|z'|^2)}
  \left( \frac{\Delta_{\{1,-1\}}[\phi_z]}{\Delta_{\{0\}}[0]}-1\right).
\label{eq:Zzcorr}
\end{equation}
} Here $f(z)$ is given by Eq.~(\ref{eq:fresult}) and we have made an
inversion transformation for the anti-instanton variable $z'\to 1/z'$.
Therefore the anti-instanton coordinate also obeys $|z|<1$.

The ratio of the Vandermonde determinants
$\frac{\Delta_{\{1,-1\}}[\phi_z]}{\Delta_{\{0\}}[0]}$ is determined by the
product over the pairs $n,m$ such that $\varepsilon_n\times \varepsilon_n<
0$, see Appendix \ref{sec:Vandermondeinnst}. We are unable to evaluate the
correlator (\ref{eq:Zzcorr}) for arbitrary values of the instanton
parameters and the gate voltage difference $q-q'$.~\cite{footnote}
However, for large differences of the gate voltage, $q-q' \gg
\Gamma/\delta$, the correlator (\ref{eq:Zzcorr}) may be evaluated with
logarithmic accuracy. Indeed, at such gate voltage differences even the
maximal possible instanton correction to the eigenvalue difference in the
denominator in Eq.~(\ref{eq:spcorr}) is relatively small, $\delta (J^i_m
-J^j_n) \sim \Gamma \ll \delta (q-q')$ (see
Eq.~(\ref{eq:eigenvalueresult})), and may be evaluated perturbatively.  In
this case $\frac{\Delta_{\{1,-1\}}[\phi_z]}{\Delta_{\{0\}}[0]}$ only
weakly depends on the instanton parameters $z$ and $z'$. Therefore the
integral is again dominated by short instantons $1-|z|^2\ll T/\Gamma$ for
which the eigenvalues $J^j_n$ are accurately described by
Eq.~(\ref{eq:eigenvalueresult}).  We show in appendix
\ref{sec:Vandermondeinnst} that for $\delta ( q -q')\gg T, \Gamma$
\begin{equation}
  \label{eq:vandcorr_result}
  1-\frac{\Delta_{\{1,-1\}}[\phi_z]}{\Delta_{\{0\}}[0]}=
  \frac{\left(\frac{4\Gamma}{\pi T}\right)^2 }{\left( b +
      \case{1}{1-|z|^2} 
  \right)\left( b + \case{1}{1-|z'|^{2}} \right) },  
\end{equation}
where $ b= i \delta ( q -q')/4 \pi T$.  

Substituting Eqs.~(\ref{eq:vandcorr_result}) and (\ref{eq:fresult}) into
(\ref{eq:spcorr}) and integrating over $z$ and $z'$ we obtain with
logarithmic accuracy the long range asymptotics of the oscillatory part of
the irreducible correlator of the thermodynamic potentials \wide{m}{
\begin{equation}
  \label{eq:connectcorrresult}
  \langle\langle\Omega(q)\Omega(q')\rangle\rangle_{\rm osc} = 
\frac{g_T^4 E_C^2}{2\pi^2 } \exp(-g_T)
\left(\frac{16 \Gamma}{\delta(q-q')}\right)^2
\ln^2\left(\frac{\delta(q-q')}{\Gamma+\pi 
      T}\right) \cos[2\pi  (q-q') ].
\end{equation}
}
This expression represents the main result of this section.

\section{Summary and discussion}
\label{sec:conclusion}

We studied the effects of Coulomb interaction on the statistics of the
thermodynamic quantities in an ensemble of weakly disordered metallic
grains with broken time reversal symmetry.  We assumed that the Thouless
energy, $E_T$, in the grain exceeds the charging energy, $E_C$, and
considered the case when the grain is connected to a metallic lead by a
tunneling contact with a large conductance $g_T \gg 1$. We found
expressions for the oscillatory parts of the average thermodynamic
potential $\langle \Omega(q) \rangle_{\rm osc}$, Eq.~(\ref{2}) and of the
correlation function $\langle \Omega(q) \Omega(q') \rangle_{\rm osc}$ at
$\delta ( q -q')\gg T, \Gamma$, Eq.~(\ref{eq:connectcorrresult}).  The
results (\ref{2}) and (\ref{eq:connectcorrresult}), apply in the
temperature range $\delta \ll \Gamma \ll T$.  For $\Gamma\ll T$ equation
(\ref{2}) coincides with the single-instanton contribution of
Ref.~\onlinecite{Grabert}. In deriving these results we have neglected the
multi-instanton contributions. This approximation is valid for $T\gg g_T^2
E_C \exp(-g_T/2)$.  The correlator of mesoscopic fluctuations,
Eq.~(\ref{eq:connectcorrresult}), decays as $(q-q')^{-2}\ln^2(q-q)$ at
$\delta |q-q'| \gg T,\Gamma$. Eqs.~(\ref{2}) and
(\ref{eq:connectcorrresult}) represent the main results of the paper.

The expressions for the oscillatory parts of the average differential
capacitance $\langle \delta C (q)\rangle$ and of its irreducible
correlator $\langle \delta C (q)\delta C (q')\rangle_{\rm osc}$ may be
easily obtained from Eqs.~(\ref{2}) and (\ref{eq:connectcorrresult}) with
the aid of Eq.~(\ref{eq:capacitdef}).

To derive our main results we employed a method based on the extension of
the Itzykson-Zuber integral~\cite{Harish-Chandra57,Zuber} to the infinite
dimensional unitary group.  In this method the $Q$-matrix degrees of
freedom (which describe the mesoscopic fluctuations) are integrated out
exactly.  As in the case $\Gamma \to 0$, at $T \gg \delta$ the replicated
partition function and its correlator can be written over the topological
classes of the phase $\phi$ which decouples the Coulomb interaction term,
see Eqs.~(\ref{eq:windingsum}) and (\ref{eq:spcorr}).  Each topological
class is characterized by the winding numbers in different
replicas~\cite{Kamenev}.  At large conductance, $g \gg 1$ the functional
integral over the phase $\phi$ within each topological class can be taken
by the saddle point method. The saddle points configurations of the phase
are characterized by the complex instanton parameters $z$.\cite{Nazarov}
At finite escape rate from the dot $\Gamma$ the functional integral over
the massive fluctuations $\tilde \phi$ around the instanton configurations
depends on the instanton length $\sim (1-|z|^2)/T$, see
Eq.~(\ref{eq:fresult}). This is in contrast to the $\Gamma = 0$
case,\cite{kor,Grabert} where it is independent of the instanton
length. This leads to the fact that the logarithmic divergence of the
integral over the zero modes of the instantons at low temperatures is cut
off at the escape rate $\Gamma$, see Eq.~(\ref{2}). The study of
mesoscopic fluctuations of Coulomb blockade oscillations in this approach
amounts to evaluating the Vandermonde determinants (\ref{eq:Vanddef}) on
the instanton configurations of the phase (\ref{eq:inst}).  In the case of
the average partition function they are equal to unity when the number of
replicas is taken to zero.  In the case of the correlator of the thermodynamic
potentials the Vandermonde determinant differs from unity and depends on
the length of the instantons.  The dependence of the Vandermonde
determinant on the instanton parameters $z$ and $z'$ results in the presence
of the logarithmic factor $\ln^2 \case{\delta (q-q')}{ \Gamma +\pi T}$ in
Eq.~(\ref{eq:connectcorrresult}).  In the $\sigma$-model formulation the
mesoscopic fluctuations are described by diffusons.  The dependence of the
diffusons on the instanton parameters is determined by the $z$-dependent
eigenvalues of the operator $\hat{J}$ given by the solutions of
Eq.~(\ref{sum}) and is easily generalizable to time-reversal invariant
non-zero-dimensional systems.\cite{AB} The method used here may have
useful applications to the $\sigma$-model treatment of interaction effects
in granulated disordered metals.\cite{Beloborodov}

\acknowledgements

We are thankful to K.~Efetov for stimulating discussions at the early
stage of this work, and to I.~Aleiner, A.~Altland, L.~Glazman, A.~Kamenev,
A.~Larkin, K.~Matveev, E.~Mishchenko and Yu.~Nazarov for valuable
discussions.  We gratefully acknowledge the support of the
Graduiertenkolleg 384 and the Sonderforschungsbereich 237 and the warm
hospitality of the Max Planck Institute in Dresden where part of this work
was performed.  This research was sponsored by the Grants DMR-9984002 and
BSF-9800338, and by the A.P.  Sloan and the Packard Foundations.

\appendix

\section{Saddle point equation}
\label{temp}

In this appendix we demonstrate that the pre-exponential factor in
Eq.~(\ref{eq:ZQreg}) gives rise only to a small correction to the saddle
point equation, Eq.~(\ref{eq:speq}) and may be neglected.  To show this it
is convenient to rewrite the integrand in Eq.~(\ref{eq:ZQreg}) in the
following form \wide{m}{ \renewcommand{\thesection}{\Alph{section}}
  \renewcommand{\theequation}{\thesection\arabic{equation}}
\begin{equation}
\exp \left\{-N \sum\limits_{m,i}F(\lambda(x_{m,i}), x_{m,i}) +
\frac{1}{2}\sum\limits_{i,j;n,m}
\ln\left(\frac{\lambda(x_{m,i})-
\lambda(x_{n,j})}{x_{m,i}-x_{n,j}}\right)\right\}.
\label{98}
\end{equation}
}

Taking a derivative of the function in the exponent in Eq. (\ref{98}) with
respect to $\lambda(x_{m,i})$ we obtain the saddle point equation for
$\lambda$
\begin{eqnarray}
&&-N\left(-\frac{1}{\lambda(x_{n,i})}+\lambda(x_{n,i})-
x_{n,i}\right) \nonumber \\
&&+\frac{\alpha}{2}\sum\limits_{m=-\infty}^{\infty}
\frac{1}{\lambda(x_{n,i})-\lambda(x_{m,i})}=0.
\label{saddle7}
\end{eqnarray}
The saddle point value for $\lambda$, Eq.~(\ref{lambda}) has been obtained
under the assumption that the second term in Eq.~(\ref{saddle7}) can be
neglected. To check the self-consistency of this assumption we now
evaluate the second term in Eq.~(\ref{saddle7}) at the saddle point
$\lambda=\lambda^0(x_{m,i})$, where $\lambda^0$ is defined by
Eq.~(\ref{lambda}).  To this end we rewrite the sum in Eq.~(\ref{saddle7})
in terms of contour integral
\begin{eqnarray}
&&\sum\limits_{m=-\infty}^{\infty}
\frac{1}{\lambda^0(x_{n,i})-\lambda^0(x_{m,j})} \nonumber \\
&& = \frac{1}{4\pi i T}\lim_{\eta \rightarrow 0}\oint\limits_C d\varepsilon
e^{i\varepsilon\eta}\left(\tan\left(\frac{\varepsilon}{2T}
\right)+i\right)\frac{1}{\lambda^0(x_{n,i})-\lambda^0(\varepsilon)},
\label{contoura}
\end{eqnarray}
where the contour $C$ is shown in Fig.~\ref{fig1}, and
$\lambda^0(\varepsilon)=\varepsilon/2+\sqrt{1+\varepsilon^2/4}$.  Taking
into account the analytic properties of $\lambda(\varepsilon)$ we can
deform the integration contour to $C'$. After a straightforward
calculation we obtain \wide{m}{
\begin{equation}
\sum\limits_{m=-\infty}^{\infty}
\frac{1}{\lambda^0(x_{n,i})-\lambda^0(x_{m,j})}=
-\frac{1}{2\pi i T\lambda^0(x_{n,i})}
\int\limits_{-2}^2 dx\left(\tanh\frac{x}{2T}+1\right)
\frac{\sqrt{1+x^2/4}}{x+i\frac{\lambda^0(x_{n,i})^2-1}
{\lambda^0(x_{n,i})}}\sim
\frac{1}{T}\ln T.
\label{45}
\end{equation}
} To obtain the last expression in Eq.~(\ref{45}) we used the fact that
$\lambda^0(x_{n,i})-1\ll 1$.

A direct comparison shows that the second term in Eq.~(\ref{saddle7}),
given by Eq.~(\ref{45}), is smaller than the first one in the ratio
$\frac{\alpha \delta}{T}\ln T$ (to obtain this estimate we used the fact
that $N=\pi/\delta$).  Therefore we may neglect the second term in
Eq.~(\ref{saddle7}) and obtain the saddle point value for $\lambda$ given
by Eq.~(\ref{lambda}).

\section{Eigenvalues of $\hat{J}$ on the instanton configuration}
\label{sec:eigenJ}

To find the eigenvalues of the operator $\hat{J}$ it is convenient to
express the operator $\hat{J}$, Eq.~(\ref{eq:J}), in terms of the complex
variable $u=\exp(i2\pi T\tau)$. Introducing the notation
$\mu_q=\case{\delta}{2}[q-N_0(\mu)]$  we write
\begin{equation}
\label{Jcomplex}
\frac{\hat{J}+i\mu_q}{2\pi T}=-u\partial_u-\frac{i\Gamma}{2\pi^2T}
\oint\frac{du'}{u'}\frac{1-uz^*}{u-z}\frac{\sqrt{uu'}}{u-u'}
\frac{u'-z}{1-u'z^*}.
\end{equation}
In Eq.~(\ref{Jcomplex}) we used that the matrices $\Lambda_{\tau,\tau'}$
may be written using the complex variable $u$ as follows:
$\Lambda=\frac{2\sqrt{uu'}}{u-u'}$.  The integral in Eq.~(\ref{Jcomplex})
is taken over the unit circle, $|u'|=1$, and the operator should be
understood as acting in the space of functions spanned by the fermionic
Matsubara frequencies $u^{-n-1/2}$.

For $|z|<1$ the matrix elements of the $\hat{J}$ in the Matsubara basis
are given by
\begin{eqnarray}
 J_{nm}&=& \left(\varepsilon_n - i\mu_q + \Gamma {\rm sgn}\varepsilon_n
\right)\delta_{nm}    \nonumber \\
&&-2 \Gamma (1-|z|^2)z^n z^{*m} \theta ( \varepsilon_n) \theta (
\varepsilon_m), 
  \label{eq:Jelements}
\end{eqnarray}
with $\theta (x)$ being the step function. It is clear that the functions
$u^{-n-1/2}$ with negative Matsubara frequencies, $n<0$, are
eigenfunctions of the operator (\ref{Jcomplex}) with eigenvalues $J_n
=\varepsilon_n-\Gamma - i\mu_q$.
  
  For $n\geq 0$ the eigenfunctions of $\hat{J}$ may be written as
\begin{equation}
  \label{eq:matsubshift}
  \frac{g^{(n)}(u)}{\sqrt{u}}=\sum\limits_{m\geq
    0}g_m^{(n)}\frac{u^{-m}}{\sqrt{u}}, 
\end{equation}
where $g^{(n)}(u)$ is a non-singular function of $u$ outside the unit
circle, $|u|>1$.  Using this property and acting with the operator
$\hat{J}$, Eq.~(\ref{Jcomplex}) on (\ref{eq:matsubshift}) we can write the
eigenvalue equation as
\begin{equation}
  \label{eq:eigeneq}
 \left[\case12 -u\partial_u +\case{\Gamma -i \mu_q - J_n (z)}{2 \pi
   T}\right]g^{(n)}(u) =
\case{\Gamma (1-|z|^2)}{ \pi
  T(1-u^{-1}z)}g^{(n)}\left(\case{1}{z^{*}}\right). 
\end{equation}
Next we introduce the Green's function of the operator on the left hand
side of this equation,
\begin{equation}
  \label{eq:greenfunc}
 G^{(n)} (u,u') =
\sum_{m\geq 0}\frac{u^{-m}u'^{m}}{J_n (z) + i \mu_q
  -\varepsilon_m- \Gamma},
\end{equation}
and rewrite Eq.~(\ref{eq:eigeneq}) as $g(u)= \case{1}{2\pi i}\oint
\case{du'}{u'} G (u,u') f(u')$, where $f(u')$ is the right hand side of
Eq.~(\ref{eq:eigeneq}). Then setting $u=1/z^{*}$ we obtain the eigenvalue
equation in the form
\begin{equation}
\label{sum}
\frac{1}{1-|z|^2}=-\sum\limits_{k=0}^{\infty}
\frac{2\Gamma |z|^{2k}}{J_n (z) +i \mu_q -\varepsilon_k -\Gamma}.
\end{equation}

In the $z\to 0$ limit we obtain that the $n=0$ eigenvalue is given by $J_0
(z\to 0)=\varepsilon_0 - i\mu_q-\Gamma$ whereas all the other eigenvalues
are unchanged: $J_n (z\to 0)=\varepsilon_n - i\mu_q + \Gamma$. In the
opposite limit of very short instantons, $1-|z|^2 \ll \Gamma/T$ (as well
as for arbitrary $z$ at $\Gamma \ll T$ ) the sum in Eq.~(\ref{sum}) is
dominated by a single term in which the denominator is small. We then
obtain for the eigenvalues with positive $\varepsilon_n$,
\begin{equation}
  \label{eq:eigenvalueresult}
 J_n (z) =\varepsilon_n+\Gamma -
i\mu_q -2\Gamma |z|^{2n}(1-|z|^2).
\end{equation}

\section{Evaluation of the Vandermonde determinant on the instanton
  configurations}
\label{sec:Vandermondeinnst}

In this appendix we calculate the Vandermonde determinant
$\Delta_{\{W\}}[\phi]$, 
Eq.~(\ref{eq:Vanddef}) on the instanton configuration of the fields
$\phi$, Eq.~(\ref{eq:inst}).

We start with the case of the average thermodynamic potential, when the
instanton with a unit winding number is present only in one replica,
$j=1$. The correction $\delta J_{n,1}$ to the eigenvalues of $\hat{J}$ in
the $j=1$ replica due to the presence of the instanton,
Eq.~(\ref{eq:eigenvalueresult}) is small in comparison with the band width
(which is of order unity in our notations). Therefore we may write the
correction to the Vandermonde determinant Eq.~(\ref{eq:Vanddef}), due to
the presence of the instanton as
\begin{equation}
  \label{eq:deltacorrpert}
  \delta\ln \Delta=4 \sum\limits_{j,n,m}
\left(\frac{\partial \lambda_0(J^0_{n,1})/\partial J^0_{n,1}}
{\lambda_0(J^0_{n,1})-\lambda_0(J^0_{m,j})}
-\frac{1}{J^0_{n,1}-J^0_{m,j}}\right)
\delta J_{n,1},
\end{equation}
where $\lambda_0(J^0_{m,j})$ is given by Eq.~(\ref{lambda5}). 
The terms with
$\varepsilon_m,\varepsilon_n > 0 $ and with $\varepsilon_m,\varepsilon_n <
0$ in the right hand side of this equation vanish. Therefore, to find the
dependence of the Vandermonde determinant on the instanton parameters $z$
we may restrict the product over $n$ and $m$ in Eq.~(\ref{eq:Vanddef})
only to the terms with $\varepsilon_m \varepsilon_n <0$. 

To evaluate the logarithm of the Vandermonde determinant 
in Eq.~(\ref{eq:Vanddef}) in the $\alpha \to 0$ limit we can use 
the following considerations.  For a given pair of Matsubara frequencies
$n$ and $m$ in the sum over the replica indices $i$ and $j$ there are: i)
one term  where both eigenvalues $J^i_n$ and $J^j_m$ of the operator
$\hat{J}$ depend on the instanton parameter $z$; ii) $2(\alpha -1)$ 
terms where one eigenvalue depends on  $z$ and the other one
does not; iii)  $(\alpha -1)^2$ terms where both eigenvalues are
independent of $z$ and equal 
to $J_n^0= \varepsilon_n + \Gamma {\rm sgn} \varepsilon_n$ and $J_m^0$. 
Thus in the $\alpha \to 0$ limit we obtain 
\begin{equation}
  \label{eq:Vandz}
 \ln \Delta= -4\sum'\limits_{n,m}\ln\left\{\frac{[J_n(z)-J_m(z)]
(J_n^0-J_m^0)}{[J_n^0-J_m(z)][J_n(z)-J_m^0]}\right\},
\end{equation}
where the summation goes over $n\geq 0$, $m<0$, $J_n^0= \varepsilon_n +
\Gamma {\rm sgn} \varepsilon_n $ are the eigenvalues of $\hat{J}$ in the
absence of instantons, and $J_n(z)$ are given by
Eq.~(\ref{eq:eigenvalueresult}).  By observing that for negative Matsubara
frequencies $J_m(z)=J_m^0$ it is easy to see that the right hand side of
this equation vanishes and $\Delta_1[\phi_z] = 1$.

Next we turn to evaluating the Vandermonde determinant for the correlator
of thermodynamic potentials, i.e. the last product in
Eq.~(\ref{eq:spcorr}).  In the limit $\alpha,\alpha' \to 0$ it can be done 
using the considerations employed above. In the case when all the winding
numbers $W_j$ have the same sign it is not difficult to see that the
Vandermonde determinant  is equal to unity.  In the case when
the instantons with both positive and negative winding numbers are present
this is no longer the case.  For the case when one instanton with a
positive winding number, $|z|<1$ and one instanton with a negative winding
number, $|\zeta|>1$ are present in the limit $\alpha,\alpha'\to 0$ we obtain
\begin{equation}
\label{detinst}
\frac{\Delta_{\{0\}}[0]}{\Delta_{\{1,-1\}}[\phi_z]}=
\prod\limits_{n,m}\left\{\frac{[J_n(z)-J_m(\zeta)]
(J_n^0-J_m^0)}{[J_n^0-J_m(\zeta)][J_n(z)-J_m^0]}\right\}^4.
\end{equation}
For $\Gamma \ll T$ the eigenvalues $J_n(z)$ for $|z|<1$ and $n \geq 0$ are
given by Eq.~(\ref{eq:eigenvalueresult}).  The expression for $J_m(\zeta)$
for $|\zeta|>1$ and $m < 0$ is given by
$J_m(\zeta)=\varepsilon_m-\Gamma - i\mu_q' +2\Gamma
|\zeta|^{2m}(|\zeta|^2-1)$. Substituting these eigenvalues into
Eq.~(\ref{detinst}) and introducing the new variable $z'=1/\zeta$ we
obtain the following expression for the exponential factor in
Eq.~(\ref{eq:spcorr})
\begin{equation}
  \label{eq:expfactor}
\frac{\Delta_{\{0\}}[0]}{\Delta_{\{1,-1\}}[\phi_z]}=
e^{\sum\limits_{n,m\geq 0}
\frac{4 |z|^{2n}|z'|^{2m}(1-|z|^2)(1-|z'|^{2})}{
\left[\case{\pi T}{\Gamma}(n+m+1)+i(q-q')\case{\delta}{4\Gamma}
  +1\right]^2}}. 
\end{equation}
The sum in the exponent can be evaluated in the limits when the lengths of
the instantons $(1-|z|^2)/T$ and $(1-|z'|^2)/T$ are either shorter or
longer than the inverse chemical potential difference $1/\delta(q-q')$. To
evaluate the correlator (\ref{eq:connectcorrresult}) with logarithmic
accuracy we can interpolate between the two limits and obtain
Eq.~(\ref{eq:vandcorr_result}).

\section{Integration over the massive modes}
\label{fluct}

In this appendix we evaluate the ratio of the integrals over the massive
fluctuations of $\phi$ around the instanton configuration in
Eq.~(\ref{zz}). The massive modes may be integrated out in the Gaussian
approximation.

We start with the limit of very long instantons, $|z| \to 0$, or $|z|\to
\infty$. In this case it is convenient to work in the Fourier basis. We
introduce the Fourier components $\phi_k$ of ${\phi}(\tau)$ as
\begin{equation}
\label{eq:phik}
\dot{\phi}(\tau)=
-i\sum\limits_{k\neq 0}\omega_k e^{-i\omega_k\tau}\phi_k,
\hspace{1cm} \omega_k=2\pi T k,
\end{equation}
and expand the eigenvalues $J^j_m$ of the operator $\hat{J}$,
Eq.~(\ref{eq:J}), to second order of perturbation theory in $\phi_k$,
\begin{equation}
\delta J^{(2)}_{n,j}=\sum\limits_{k\neq
0}\frac{|\dot{\phi}^j_{k}|^2}{\varepsilon_n-\varepsilon_{n-k}
+\Gamma(\rm sgn\varepsilon_n-\rm sgn \varepsilon_{n-k})}.
\label{correcJ}
\end{equation}
Expanding the free energy ${\cal F}^j_{\{ W\}} (q,\phi)$ given by
Eqs.~(\ref{free5'}) and (\ref{eq:Fab}) in $\delta J^{(2)}_{n,j}$,
\wide{m}{
\begin{equation}
\delta{\cal F}^j_{\{ W \}}= 2NT\sum\limits_{n}\left(
\frac{\partial F(\lambda_0(J_{n,j}),J_{n,j})}
{\partial\lambda_0(J_{n,j})}\frac{\partial\lambda_0(J_{n,j})}
{\partial J_{n,j}}
+\frac{\partial F(\lambda_0(J_{n,j}),J_{n,j})}
{\partial J_{n,j}}\right)\delta J^{(2)}_{n,j}
=-2TN\sum\limits_{n}\left(\lambda_0(J^0_{n,j})-J_{n,j}^0\right)\delta
J_{n,j}^{(2)}. 
\label{free44}
\end{equation}
} To arrive at Eq.~(\ref{free44}) we used the fact that $\partial
F(\lambda,J_{n,j})/\partial\lambda=0$ at the saddle point.  Substituting
Eq.~(\ref{correcJ}) into Eq.~(\ref{free44}) we obtain for the second order
correction to the free energy
\begin{equation}
\delta{\cal F}^j_{\{ W \}}=
-2NT\sum\limits_{n,m} \frac{|\dot{\phi}^j_{n-m}|^2\left({
      \lambda}_0(J^0_{n,j})- 
J_{n,j}^0\right)}{\omega_{n-m}
+\Gamma(\rm sgn\varepsilon_n-\rm sgn \varepsilon_m)},
\label{7}
\end{equation}
In Eq.~(\ref{7}) the summation over $n$ and $ m$ goes over $n \neq m$.
Next we collect the terms with $|\dot{\phi}^j_k|^2$, $k\neq 0$. To this
end we retain only the terms with $m=n \pm k$ in the sum over $m$ in the
right hand side of Eq.~(\ref{7}).  As a result we obtain
\begin{equation}
\delta{\cal F}^j_{\{ W \}}=-2NT\sum\limits_{
n}|\dot{\phi}^j_{k}|^2\left(
\lambda_0(J^0_{n,j})-J_{n,j}^0\right)X_{n,k},
\label{14}
\end{equation}
where we introduced the notation
\begin{equation}
X_{n,k}=\case{1}{2\pi Tk
+\Gamma(\rm sgn\varepsilon_n-\rm sgn \varepsilon_{n-k})} +
\case{1}{-2\pi Tk
+\Gamma(\rm sgn\varepsilon_n-\rm sgn \varepsilon_{n+k})}.
\label{13}
\end{equation}
The function $X_{n,k}$ vanishes for $|\varepsilon_n|>2\pi Tk$.  Therefore
the sum over $n$ in Eq.~(\ref{14}) is restricted to $|\varepsilon_n|< 2\pi
Tk$ and can be straightforwardly evaluated leading to the following second
order correction to the free energy
\begin{equation}
\delta{\cal F}^j_{\{ W \}}=
2 \sum\limits_{k\geq 1}\frac{(2\pi T)^2\Gamma |k|
\left(|k|-|W_j|\right)}{\delta (\pi T
|k| +\Gamma ) } |\phi_k^j|^2 \theta(|k|-|W_j|).
\label{eq:deltaFresult}
\end{equation}

Using Eq.~(\ref{eq:deltaFresult}) we obtain for the ratio of the integrals
over the Fourier components $\phi_k$ with $k>1$ in Eq.~(\ref{zz}),
\wide{m}{
\begin{equation}
\left(\frac{2\pi^2T}{E_C}+
\frac{\pi T g_T }{\pi T  +\Gamma }\right)
\prod\limits_{k=2}^{\infty}\left(\frac{
\frac{2\pi^2Tk^2}{E_C}+
\frac{\pi T g_Tk^2 }{\pi T
|k| +\Gamma }}
{\frac{2\pi^2Tk^2}{E_C}+\frac{\pi T g_T|k|
  \left(|k|-1\right)}{\pi T
|k| +\Gamma } }\right)
=\frac{g_T^2 E_C}{2 \pi(\pi T+\Gamma)}.
\label{99}
\end{equation}
} In the limit $\Gamma \ll T$, Eq.~(\ref{99}) reproduces the result of
Ref.~\onlinecite{Grabert}.

Next we evaluate the ratio of the integrals over the massive fluctuations
of $\phi$ in Eq.~(\ref{zz}) in the limit of short instantons, $1-|z|^2 \ll
T/\Gamma$.  To this end we use Eq.~(\ref{free44}) to rewrite the
correction to the free energy in the form
\begin{equation}
\label{short}
\delta F=-NT\sum_{n,\sigma}\rm sgn \varepsilon_n \delta J_n,
\end{equation}
where $\delta J_n$ up to second order in $\phi$ is given by
\begin{equation}
\label{J2}
\delta J_n=\sum_{m\neq n}\frac{\left|  \langle |\varphi_n |
    \dot{\phi}|\varphi_m\rangle\right|^2}{J_n(z)-J_m(z)}, 
\end{equation}
here $ |\varphi_n\rangle$ are the eigenfunctions of the operator $\hat{J}$
and $J_n(z)$ are its eigenvalues.  For the positive frequencies
$\varepsilon_n$ the latter are given by Eq.~(\ref{eq:eigenvalueresult}).

For a finite length of the instanton the operator $\hat{J}$ is no longer
diagonal in the Fourier basis.  However, in the limit of short instantons,
$1-|z|^2 \ll T/\Gamma$, the non-diagonal matrix elements of $\hat{J}$,
Eq.~(\ref{eq:Jelements}), are small and may be treated perturbatively.
Treating the instanton correction to $\hat{J}$ given by the last term in
Eq.~(\ref{eq:Jelements}), $\gamma_{m,n}=-2\Gamma (z^*)^n
z^m(1-|z|^2)\theta(\varepsilon_n) \theta(\varepsilon_m)$ as a perturbation
we can write the eigenfunctions in Eq.~(\ref{J2}) as
\begin{equation}
\label{perturbation}
|\varphi_n\rangle =|n \rangle + \sum_{m \neq n}\frac{| m \rangle
  \gamma_{mn}}{J^0_n-J^0_m}, 
\end{equation}
here $J^0_n=\varepsilon_n+\Gamma \rm sgn \varepsilon_n$ and $|n \rangle $
are the eigenvalues and the eigenfunctions of the operator $\hat{J}$ in
the absence of the instanton.  Substituting Eq.~(\ref{perturbation}) into
Eq.~(\ref{J2}) and expanding the denominator Eq.~(\ref{J2}) to linear
order in $\gamma_{n,n}$ after a lengthy but straightforward calculation we
obtain the correction to the free energy in Eq.~(\ref{short}) in the form
$\delta F= \frac{T g_T}{2}\sum_{l,l'}\phi_l^*\phi_{l'} \delta F_{l,l'}$, where 
\wide{m}{
\begin{equation}
\label{matsubara}
\delta
F_{l,l'} =
  \frac{\delta_{l,l'} l^2}{|l|+\Gamma/\pi T}+ 
\frac{|l||l'| \left (z^{|l|+|l'|}-z^{\left | |l|-|l'| \right |}\right
  )}{\left( |l| +\Gamma/\pi T\right) 
\left( |l'| +\Gamma/\pi T\right)  } \left[
\theta(l)\theta(l')+\theta(l)\theta(l')\right]. 
\end{equation}
} Since the functional integral over the fluctuations depends only on the
absolute value of $z$ we assumed $z$ to be real without loss of
generality.
Equation (\ref{matsubara}) is written in the Matsubara basis.  It is more
convenient to use the basis of Ref.~\onlinecite{kor} which diagonalizes
the matrix $\delta \hat{F}$, Eq.~(\ref{matsubara}), at
$\Gamma=0$.  Denoting the functions in this basis by $|
\tilde{\varphi}_m\rangle$, where
\begin{equation}
| \tilde{\varphi}_m \rangle= \left\{
\begin{array}{lr}
\frac{\sqrt{1- |z|^2}}{u-z}, & m =1\\
u^{-m}\frac{1-uz^*}{1-u^{-1}z}, & m \geq 2 \\
u^{-m}\frac{1-u^{-1}z}{1-uz^*}, & m \leq -2 \\
\frac{\sqrt{1- |z|^2}}{u^{-1}-z^*}, & m =-1,
\end{array}
\right.
\end{equation}  
we can express the matrix elements of $\delta \hat{F}$ in this basis as 
\begin{equation}
\label{basis}
\delta F_{nm}=
\sum_{l,l'}\langle \tilde{\varphi_n} | l \rangle \delta F_{l,l'} \langle
l' | \tilde{\varphi}_m\rangle, 
\end{equation}
where $\langle l | \tilde{\varphi}_m \rangle$ are the elements of the
unitary transformation matrix between the two bases,
\begin{equation}
\label{matrix}
\langle l  | \tilde{\varphi}_m \rangle= \left\{
\begin{array}{lr}
z^{l-m}\left( 1- |z|^2 \right), & l \geq m \\
-z^*, & l=m-1 \\
z^{l-1}\sqrt{1-|z|^2}, & m=1.
\end{array}
\right.
\end{equation}
Substituting Eqs.~(\ref{matrix}), (\ref{matsubara}) into Eq.~(\ref{basis})
after 
straightforward calculations in the limit of short instantons, $1-|z|^2
\ll T/\Gamma$, for the matrix elements $\delta F_{nm}$ in the new basis we
obtain
\begin{equation}
\label{newbasis}
\frac{\delta F_{nm}}{T}=\frac{g_T}{2}\left[
  \frac{(n-1)^2\delta_{n,m}}{n-1+a}+ 
\frac{a^2\left( z^{n+m}-z^{n-m}\right)}{(n-1+a)(m-1+a)}\right]. 
\end{equation}
Here $n,m \geq 2$ and $a=\Gamma/\pi T$.  For $\Gamma =0$ the matrix
$\delta \hat{F}$ in Eq.~(\ref{newbasis}) becomes diagonal and independent
of the instanton size, in agreement with Ref.~\onlinecite{kor}. At finite
$\Gamma$ and in the limit of short instantons, $1-|z|^2 \ll T/\Gamma$, the
off-diagonal matrix elements in (\ref{newbasis}) are small and may be
neglected for the purpose of evaluating the ratio of the two integrals
over the fields $\tilde \phi$ in Eq.~(\ref{eq:fdef}).  Proceeding this way
we obtain the following expression for the ratio of the two integrals over
the massive modes in Eq.~(\ref{eq:fdef}) for $1-|z|^2 \ll T/\Gamma$
\begin{equation}
\label{detshort}
\frac{g_T^2E_C}{2\pi^2 T}\left[1-\frac{\Gamma}{\pi
    T}(1-z^2)\ln\left(1+\frac{\Gamma}{\pi T}\right)\right]. 
\end{equation}
Note that in the limit $z \to 1$ this expression does not contain $\Gamma$
and coincides with the result of Ref.~\cite{Grabert}.

\end{multicols}

\begin{references}
\bibitem{footnote} For $T\leq \Gamma/\pi T$ and at $q=q'$ the eigenvalues
  $J^i_n$ and $J^j_m$ can become degenerate at certain values of the
  instanton parameters $z$ and $z'$.  At these points
  $\frac{\Delta_{\{1,-1\}}[\phi_z]}{\Delta_{\{0\}}[0]}$ develops
  singularities which dominate the integral in Eq.~(\ref{eq:Zzcorr}).
\end{references}
\end{document}